\documentclass[fleqn,10pt]{wlscirep}
\usepackage[utf8]{inputenc}
\usepackage[T1]{fontenc}

\title{Nonlinear methods to quantify Movement Variability in Human-Humanoid Interaction Activities}
\author[1,*]{Miguel Xochicale}
\author{Chris Baber}
\affil[1]{University of Birmingham,
	School of Computer Science,
	Birmingham, 
	B15 2TT, 
	UK}
\affil[*]{perez.xochicale@gmail.com}

\begin{abstract}
Human movement variability arises from the process of mastering redundant (bio)mechanical degrees of freedom to successfully accomplish any given motor task where flexibility and stability of many possible joint combinations helps to adapt to environment conditions.
While the analysis of movement of variability is becoming increasingly popular as a diagnostic tool or skill performance evaluation, there are remain challenges on applying the most appropriate methods.
We therefore investigate nonlinear methods such as reconstructed state space (RSSs), uniform time-delay embedding, recurrence plots (RPs) and recurrence quantification analysis (RQAs) with real-world time-series data of wearable inertial sensors.
That said, twenty healthy participants imitated vertical and horizontal arm movements in normal and faster velocity from an humanoid robot.
We applied nonlinear methods to the collected data to found visual differences in the patterns of RSSs and RPs and statistical differences with RQAs.
We conclude that Shannon Entropy with RQA is a robust method that helps to quantify activities, types of sensors, windows lengths and level of smoothness.
Hence this work might enhance the development of better diagnostic tools for applications in rehabilitation and sport science for skill performance or new forms of human-humanoid interaction for quantification of movement adaptations and motor pathologies.
\end{abstract}

\begin{document}
\flushbottom
\maketitle
\thispagestyle{empty}

\section*{Introduction}
The complexity of human movement arises from the process of mastering redundant (bio)mechanical degrees of freedom (DOF) to successfully accomplish any given motor task (Bernstein, 1967).
Such DOF are independent coordinates to uniquely describe the configurations of the system which provides flexibility and stability to adapt to a change environment conditions but that leads many possible combinations (Newell and Vaillancourt, 2001).
Hence, human movement complexity can be seen as a balance between the required DOF "to generate a stable (persistent) and flexible (variable) behavioural output in response to changing intentions and dynamic environmental conditions" \cite{davids2003}.
Consequently, one can see much variability in human movement even in the simplest of movements.
While the analysis of movement of variability is becoming increasingly popular as a diagnostic tool or skill performance evaluation, there remain challenges in terms of defining the most appropriate methods and parameters to apply.
In part, these challenges stem from the fact that the identification movement variability requires analysis of signals which are time-series of $1-$dimension in $\mathbb{R}$ which are noisy, nonlinear and non-stationary \cite{gomezgarcia2014}.
Further problems arise from assigning a plausible locus of control to the movements, e.g., in order to determine whether variability is the result of deliberative control by the performer or whether it arises from exogenous or endogenous disturbances.
For this paper, our focus is on the analysis of signals; specifically, in terms of objectively quantifying variability of lower dimension signals using time-series analysis.

Methods for nonlinear time series analysis generally involve the estimation of the embedding parameters ($m$ embedding dimension and $\tau$ embedding delay) to reconstruct the state space, where an $n$-dimensional is reconstructed state space using $1-$dimensional time series \cite{Quintana-Duque2012, Quintana-Duque2016, sama2013, frank2010, gomezgarcia2014, marwan2011, stergiou2011}.
Key to the selection of these parameters is the need to preserve topological properties of an unknown $M$-dimensional state space \cite{takens1981}.
An common approach to the construction of these state spaces involves Recurrence Plots (RPs), a graphical representation of black and white dots, which shows recurrent patterns of a $n$-dimensional system.
While RPs provide a human interpretable picture of the system, these require further analysis to allow the properties of that system to be quantified, and so Recurrence Quantification Analysis (RQA) can be applied.
However, the estimation of the embedding parameters for RQA remains an open problem (Bradley et al. 2015 \cite{bradley2015}).
There is no agreed method to estimate embedding parameters for RQA and other nonlinear methods \cite{bradley2015} because time series are system-dependent, i.e., these rely on the initial conditions, and on the configuration and behaviour of the system.
This means that, unless one holds all of the influencing variables constant, embedding parameters computed for one instance may not apply to another instance.

One could apply methods, such as autocorrelation, mutual information, nearest neighbour (and we will apply these in this paper), but these methods assume that the data are well sampled, with little noise and (usually) that the signals are purely deterministic \cite{garland2016, kantz2003}.
That said, these methods can break-down in the face of real-world datasets which could have different length, different values of accuracy and precision (rounding errors due to finite precision of the measurement apparatus which include frequency acquisition \cite{frank2010}), or different levels of contamination from exogenous and endogenous sources of ‘noise’ \cite{garland2016}.
What is, perhaps, surprising is that even subject to these problems, the methods are useful and nonlinear dynamics approaches continue to tell us much about human movement \cite{Quintana-Duque2012, Quintana-Duque2016, sama2013, frank2010, gomezgarcia2014, marwan2011, stergiou2011, bradley2015}.

For this paper, we explore the role of RPs and RQA in the analysis of simple human movement.
We compare horizontal and vertical arm movements performed by neurotypical participants who are copying these movements being made by a humanoid robot.
This provides an initial point of comparison (between human movement, which we assume to have a degree of variability, and the robot, which we assume to have a degree of consistency).
Additionally, our interest lies in the estimation of embedding parameters in light of different features of the data, e.g., levels of smoothing, window length, structure of time-series based on movement, types of sensors, individual differences between participants and the movements that they perform.
Specifically, we ask what are the effects of different embedding parameters, recurrence thresholds and characteristics of time-series on nonlinear analysis methods (i.e., reconstructed state space with uniform time-delay embedding, RPs and RQA)?

\section*{Methods}
\subsection*{State Space Reconstruction}
The method of state space reconstruction \cite{packard1980, takens1981} has been applied in many disciplines \cite{aguirre2009, stergiou2011, frank2010, sama2013, Quintana-Duque2016}.
The method of state space reconstruction is based on uniform time-delay embedding methodology which is a simple matrix implementation that can reconstruct an unknown $d-$dimensional manifold $M$ from a scalar time series (e.g. one-dimensional time series in $\mathbb{R}$).
A manifold, in this context, is a multidimensional curved surface within a space (e.g. a saddle) \cite{guastello-gregson2011}.

The use of a scalar time series is the main advantage of the method of state space reconstruction which in essence preserve dynamic invariants such as correlation dimension, fractal dimension, Lyapunov exponents, Kolmogorov-Sinai entropy and detrended fluctuation analysis \cite{bradley2015, Quintana-Duque2012, Quintana-Duque2013, Quintana-Duque2016, krakovska2015}.
However, selecting appropriate embedding parameters which are sued to apply the state space reconstruction is still an open challenge for which we present introductions for the methodologies to compute such embedding parameters.
With that in mind, in the following subsections, we describe in more detail the state space reconstruction theorem (RSSs), uniform time-delay embedding theorem (UTDE), false nearest neighbours (FNN) and average mutual information (AMI).

\subsection*{State Space Reconstruction Theorem}
\begin{figure}[ht]
\centering
\includegraphics[width=1.0\textwidth]{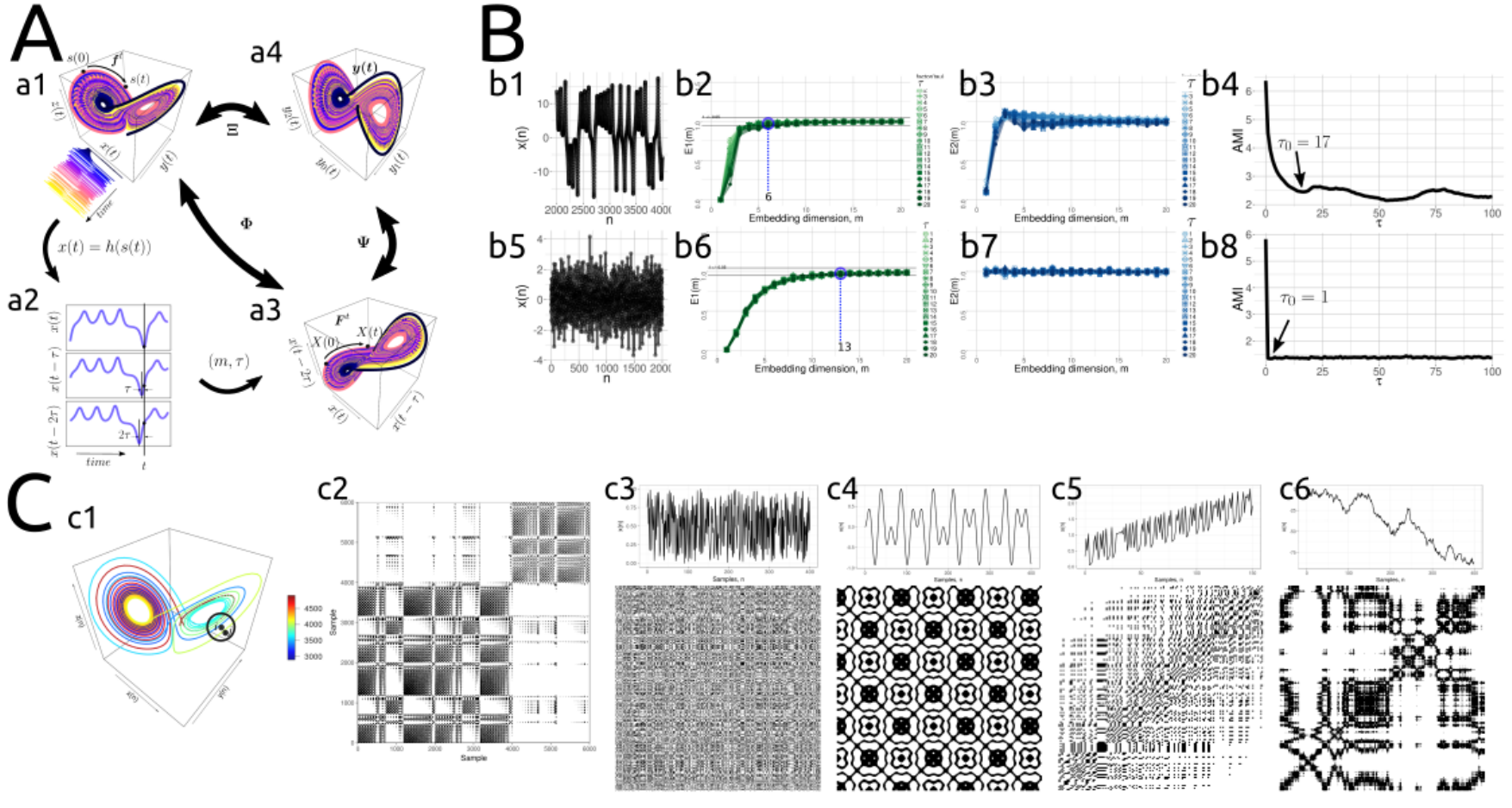}
    \caption{
	{\bf State space reconstruction methodology (A),
		Minimum delay embedding values with AMI's method and minimum dimension embedding values with Cao's method (B), and Recurrence Plots and its patterns (C).
	}
		State space reconstruction is based on $x(t)=h[s(t)]= h[f^t [s(0)]]$ where $f^t$ is the true dynamical system, $s(t)$ indicates the state, $s$, at time, $t$,  and $h[ ]$ the measurement function.
		The time-delay embedding represented as the $\Phi$, maps the original $d-$dimensional state $s(t)$ into the $m-$dimensional uniform time-delay embedding matrix $\boldsymbol{X}(t)$.
    	The transformation map $\Psi$ then maps $\boldsymbol{X}(t)$ into a new state $y(t)$ of dimensions $n < m$.
    	(a1) $M-$dimensional manifold representing the state space (e.g. Lorenz system);
    	(a2) Delayed copies of $1-$dimensional $x(t)$ from the Lorenz system;
		(a3) $m-$dimensional reconstructed state space with \texorpdfstring{$m$}{m} and    \texorpdfstring{$\tau$}{T}, and
    	(a4) $y(t)$ is the $n-$dimensional reconstructed state space.
	The total reconstruction map is represented as $\Xi = \Psi \circ \Phi $where $\Phi$ is the delay reconstruction map and $\Psi$ is the coordinate transformation map.
	a1 to a4 figures are adapted from \cite{Quintana-Duque2012, casdagli1991, uzal2011}.
	(b1) chaotic and (b5) random time series.
	(b2, b6) $E_1 (m)$ values and (b3, b7) $E_2(m)$ values with variations of $\tau$ values from one to twenty for chaotic and random time series.
	(b4, b8) AMI values where its first minimum value in the curve is the minimum time delay embedding ($\tau_0$).
	(c1) State space of the Lorenz system with controlling parameters ($\rho=28, \sigma=10, \beta=8/3$) where the point, $j$, in trajectory $X()$ which falls into the neighborhood (black circle) of a given point at $i$ is a recurrent point and is represented as a black dot in the recurrence plot at location $(i, j)$ or white otherwise.
	(c2) Recurrence plot using the three components of the Lorenz system and the RP with no embeddings and threshold $\epsilon=5$.
	Time-series with its respective recurrence plot for:
	(c3) uniformly distributed noise,
	(c4) super-positionet harmonic oscillation ($sin( \frac{1}{5}*t) * sin( \frac{5}{100}*t) $),
	(c5) drift logistic map ($x_{i+1} = 4 x_i (1- x_i) $) corrupted with a linearly increase term ($0.01*i$), and
	(c6) disrupted brownian motion  ($x_{i+1} = x_i + 2*rnorm(1) $).
	Figures in C are adapted from \cite{marwan2015}.
	Code and data to reproduce the figure is available in \cite{srep2021}.
    }
    \label{fig01}
\end{figure}

Following the notation employed in \cite{casdagli1991, garland2016, gibson1992, uzal2011, uzal2010, takens1981}, state space reconstruction is defined by:
\begin{equation}\label{eq:ssr}
  s(t)=f^t [s(0)],
\end{equation}
where $s$, $s: A \rightarrow M$ given that $A \subseteq \mathbb{R}$ and $M \subseteq \mathbb{R}^d$, represents a trajectory which evolves in an unknown $d-$dimensional manifold $M$, $f: M \rightarrow M$ is an evolution function and $f^t$, with time evolution $t \in \mathbb{N}$, is the $t$-th iteration of $f$ that corresponds to an initial position $s(0) \in M $ \cite{takens1981}.
Then, a point of a one-dimensional time series $x(t)$ in $\mathbb{R}$, can be obtained with
\begin{equation}\label{eq:measurement}
  x(t)=h[s(t)],
\end{equation}
where $h$ is a function, $h: M \rightarrow \mathbb{R}$, defined on the trajectory $s(t)$.
Reconstructed state space can then be described as an $n-$dimensional state space defined by $y(t)=\Psi[\boldsymbol{X}(t)]$ where $\boldsymbol{X}(t) = \{ x(t), x(t-\tau) , ...,x(t - (m-1)\tau  ) \}$ is the uniform time-delay embedding with a dimension embedding $m$ and delay embedding $\tau$ and $ \Psi: \mathbb{R}^m \rightarrow \mathbb{R}^n$ is a further transformation of dimensionality (e.g. Principal Component Analysis,
Singular Value Decomposition, etc) being $n \leq m$.
With that in mind, uniform time-delay embedding, $\boldsymbol{X}(t)$, defines a map $\Phi: M \rightarrow \mathbb{R}^m$ such that $\boldsymbol{X}(t) = \Phi(s(t))$, where $\Phi$ is a diffeomorphic map \cite{takens1981} whenever $\tau > 0$ and $m > 2d_{box}$ and $d_{box}$ is the box-counting dimension of $M$ \cite{garland2016}.
Then, if $\Phi$ is an embedding of evolving trajectories in the reconstructed space then a composition of functions represented with $F^t$ is induced on the reconstructed state space determined:
\begin{equation}\label{eq:st}
  \boldsymbol{X}(t)=F^t [\boldsymbol{X}(0)] = \Phi \circ f^t \circ \Phi ^{-1}[\boldsymbol{X}(0)].
\end{equation}
With this in mind, an embedding is defined as "a smooth one-to-one coordinate transformation with a smooth inverse" and the total reconstruction map is defined as $ \Xi = \Psi \circ \Phi $ \cite{casdagli1991}.
Figures~\ref{fig01}A illustrates the state space reconstruction.

\subsection*{Uniform Time-Delay Embedding (UTDE)}\label{sec:utimedelayembedding}
Frank et al. and Sama et al. refer to the state space reconstruction as "time-delay embeddings" or "delay coordinates" \cite{frank2010, sama2013}.
However, we consider the term "uniform time-delay embedding" as more descriptive and appropriate terminology for our work.
Hence, the uniform time-delay embedding is represented as a matrix of uniform delayed copies of the time series $\{ \boldsymbol{x}_n \}_{n=1}^N$ where $N$ is the sample length of $\{ \boldsymbol{x}_n \}$ and $n$ is index for the samples of $\{ \boldsymbol{x}_n \}$.
$\{ \boldsymbol{x}_n \}_{n=1}^N$ has a sample rate of $T$.
The delayed copies of $\{ \boldsymbol{x}_n \}$ are uniformly separated by $\tau$ and represented as $\{\boldsymbol{ \tilde{x} }_{n- i\tau} \}$ where $i$ goes from $0,1, \dots, (m-1)$.
Generally speaking, $\{\boldsymbol{ \tilde{x} }_{n- i\tau} \}$ contains information of unobserved state variables and encapsulates the information of the delayed copies of the available time series in the uniform time-delay embedding matrix $\boldsymbol{X}^{m}_{\tau}$, $\boldsymbol{X}^{m}_{\tau} \in \mathbb{R}^m$, defined as
\begin{equation}\label{eq:tde}
\boldsymbol{X}^{m}_{\tau}  =
\begin{pmatrix}
\boldsymbol{ \tilde{x} }_n \\
\boldsymbol{ \tilde{x} }_{n-\tau} \\
\boldsymbol{ \tilde{x} }_{n-2\tau} \\
\vdots \\
\boldsymbol{ \tilde{x} }_{n- (m-1) \tau} \\
\end{pmatrix}^\intercal, 
\end{equation}
where $m$ is the embedding dimension, $\tau$ is the embedding delay and $ ^\intercal$ denotes the transpose.
$m$ and $\tau$ are known as embedding parameters.
The matrix dimension of $ \boldsymbol{X}_{\tau}^{m} $ is defined by $N-(m-1)\tau$ rows and $m$ columns and $N-(m-1)\tau$ defines the length of each delayed copy of $\{ \boldsymbol{ \tilde{x} }_n \}$ in $\boldsymbol{X}^{m}_{\tau}$.

\subsection*{Estimation of Embedding Parameters}
The estimation of the embedding parameters ($m$ and $\tau$) is a fundamental step for the state space reconstruction with the use of uniform time-delay embedding method.
With this in mind, we review two of the most common algorithms, which will be used in our work, to compute the embedding parameters: the false nearest neighbour (FNN) and the average mutual information (AMI).

\subsubsection*{False Nearest Neighbours}
To select the minimum embedding dimension $m_0$, Kennel et al. \cite{kennel1992} used the method of false neighbours which can be understood as follows:
on the one hand, when the embedding dimension is too small to unfold the attractor "not all points that lie close each other will be neighbours and some points appear as neighbours because of the attractor has been projected down into an smaller space", on the other hand, when increasing the embedding dimension "points that are near to each other in the sufficient embedding dimension should remain close as the dimension increase from $m$ to $m+1$ \cite{krakovska2015}".
From a mathematical point of view, the state space reconstruction theorem is done when the attractor is unfolded with either the minimum embedding dimension, $m_0$, or any other embedding dimension value where $m \ge m_0$ \cite{kennel1992}. On the contrary, any large value of $m_0$ leads to excessive computations \cite{bradley2015}.
With this in mind, Cao \cite{Cao1997} proposed an algorithm based on the false neighbour method where only the time-series and one delay embedding value are necessary to select the minimum embedding dimension.
Cao's algorithm is based on $E(m)$  which is the mean value of all $a(i,m)$, both defined as follows
\begin{equation}\label{eq:e}
  \begin{aligned}
	E(m) &= \frac{1}{N-m\tau} \sum_{i=1}^{N-m\tau} a(i,m) \\
	 &=
       \frac{1}{N-m\tau} \sum_{i=1}^{N-m\tau}
       \frac{ || \boldsymbol{X}_i(m+1) - \boldsymbol{X}_{n(i,m)}(m+1) || }
            { || \boldsymbol{X}_i(m) - \boldsymbol{X}_{n(i,m)}(m) ||  }
  \end{aligned}
\end{equation}
where $\boldsymbol{X}_i(m)$ and $\boldsymbol{X}_{n(i,m)}(m)$ are the time-delay embeddings with $i=1,2,\dots,N-(m-1)\tau$ and $ n(i,m)= 1 \le n(i,m) \le N-m\tau$.
From Eq.~\ref{eq:e}, it can be seen that $E(m)$ is only dependent on $m$ and $\tau$ for which $E_1(m)$ is defined as
\begin{equation}\label{eq:e1}
E_1(m) = \frac{ E(m+1) } { E(m)}.
\end{equation}
$E_1(m)$ is therefore considered to investigate the variation from $m$ to $m+1$ in order to find the minimum embedding dimension $m_0$ (Eq.~\ref{eq:e1}).
As described in \cite{Cao1997}: "$E_1(m)$ stops changing when $m$ is greater than some $m_0$, if the time series comes from a multidimensional state space then $m_0 + 1$ is the minimum dimension".
Additionally, Cao proposed $E_2(m)$ to distinguish deterministic signals from stochastic signals. $E_2(m)$ is defined as
\begin{equation}\label{eq:e2}
E_2(m) = \frac{ E^* (m+1) } { E^*(m)},
\end{equation}
where
\begin{equation}\label{eq:ee}
E^*(m) = \frac{1}{N-m\tau} \sum_{i=1}^{N-m\tau}
|  \boldsymbol{X}_i(m+1) - \boldsymbol{X}_{n(i,m)}(m+1) |.
\end{equation}
For instance, when the signal comes from random noise (values that are independent from each other), all $E_2(m)$ values are approximately equal to 1 (e.g. $E_2(m) \approx 1$). However, for deterministic data $E_2(m)$ is not constant for all $m$ (e.g. $E_2(m) \neq 1$).

As an example of the use of $E_1(m)$ and $E_2(m)$ values, we consider two time series: the solution for the $x$ variable of the Lorenz system (Fig~\ref{fig01}b1), and a Gaussian noise time series with zero mean and a variance of one (Fig~\ref{fig01}b5).
We then compute $E_1(m)$ and $E_2(m)$ values for each time series.
The $E_1(m)$ values for the chaotic time series appear to be constant after the dimension is equal to six.
The determination of six is given that any value of $m$ can be used as they are within the threshold of $1\pm0.05$ (Fig~\ref{fig01}b2).
$E_2(m)$ values, for chaotic time series, are different to one (Fig~\ref{fig01}b3), for which, it can be concluded that for the chaotic time series the minimum embedding dimension the time series comes from a deterministic signal. With regard to the noise time series, $E_1(m)$ values appeared to be constant when $m$ is close to thirteen, which is defined by the threshold of $1\pm0.05$ (Fig~\ref{fig01}b6).
$E_1(m)$ values then indicate the minimum embedding dimension of the noisy time series is thirteen, however all of the $E_2(m)$ values are approximately equal to one (Figure~\ref{fig01}b7) for which it can be concluded that noise time series is a stochastic signal.
It is important to note that for this work not only $E_1(m)$ and $E_2(m)$ are computed but also a variation of $\tau$ from 1 to 20 is presented.
The purpose of such variation for $\tau$ is to show its independence with regard to $E_1(m)$ and $E_2(m)$ values as $\tau$ is increasing (Fig~\ref{fig01}b2,b3,b6, and b7).
However, one negative of the Cao's algorithm \cite{Cao1997} is the definition of a new threshold where $m$ values appear to be constant in $E_1 (m)$.
In the case of the given examples and reported results, we defined such threshold as 0.05.
Further investigation is required for the selection of the threshold in the $E_1(m)$, as the selection of the threshold in this work is base on no particular method but visual inspection.

\subsubsection*{Average Mutual Information}
When selecting the delay dimension parameter, $\tau$, one can consider the following two cases:
(i) when $\tau$ is too small, the elements of time-delay embedding will be along the bisectrix of the phase space and the reconstruction is generally not satisfactory,
(ii) on the contraty, when $\tau$ is too large the elements of the uniform time-delay embedding will become spread and uncorrelated which makes recovering the underlying attractor more difficult if not impossible \cite{casdagli1991, emrani2014a, garcia2005e71}.
With regard to the algorithms to compute $\tau$, Emrani et al. \cite{emrani2014a}, for instance, used the autocorrelation function in which the first zero crossing is considered as the minimum delay embedding parameter.
However, the autocorrelation function is a linear statistic over which the Average Mutual Information (AMI) algorithm is preferred because the AMI takes into account the nonlinear dynamical correlations \cite{afraser1986,krakovska2015}.
With this in mind, the AMI algorithm is described below to estimate the minimum delay embedding parameter, \texorpdfstring{$\tau_0$}{T}.

To compute the AMI, an histogram of $x(n)$ using $n$ bins is calculated and then a probability distribution of data is computed \cite{kantz2003}.
AMI is therefore denoted by $I(\tau)$ which is the average mutual information between the original time series, $x(n)$, and the delayed time series, $x(n-\tau)$, delayed by $\tau$ \cite{kabiraj2012}.
AMI is defined by
\begin{equation}\label{eq:ami}
I(\tau) = \sum_{i,j}^N p_{ij} log_2 \frac{ p_{ij} }{ p_i p_j }.
\end{equation}
Probabilities are defined as follows:
$p_i$ is the probability that $x(n)$ has a value inside the $i$-th bin of the histogram, $p_j$ is the probability that $x(n+\tau)$ has a value inside the $j$-th bin of the histogram and $p_{ij}(\tau)$ the probability that $x(n)$ is in bin $i$ and $x(n+\tau)$ is in bin $j$. The AMI is measured in bits (base 2, also called shannons) \cite{kantz2003, nonlinearTseries2016}.
For small $\tau$, AMI will be large and it will then decrease more or less rapidly. As $\tau$ increase and goes to a large limit, $x(n)$ and $x(n+\tau)$ have nothing to do with each other and $p_(ij)$ is factorised as $p_ip_j$ for which AMI is close to zero.
Then, in order to obtain $\tau_0$, "it has to be found the first minimum of $I(\tau)$ where $x(n+\tau)$ adds maximal information to the knowledge from $x(n)$, or, where the redundancy is the least" \cite{kantz2003}.

For example, we compute the AMI for two time series: A) the $x$ solution of the Lorenz system, and B) a noise time series using a normal distribution with mean zero and standard deviation equal to one.
From Fig~\ref{fig01}(b4, b8), it can then be concluded that the amount of knowledge for any noise time series is zero for which the first minimum embedding parameter is $\tau_0=1$.
On the contrary, the first minimum of the AMI for the chaotic time series is $\tau_0=17$ which is the value that maximize the independence between $x(n)$ and $x(n+\tau)$ in the reconstructed state space \cite{bradley2015}.
Similarly as Cao's algorithm negatives, AMI's algorithm is not an exception for negatives, which are worthwhile to mention for further investigations.
For instance, (i) is not clear why the choose of the first minimum of the AMI is the minimum delay embedding parameter \cite{kantz2003} and (ii) the probability distribution of the AMI function is computed with the use of histograms which depends on a heuristic choice of number of bins for which AMI depends on partitioning \cite{garcia2005e71}.

\section*{Recurrence Quantification}\label{sec:recurrence-quantification}
\subsection*{Recurrence Plots}
Originally, Henri Poincar\'e in 1890 introduced the concept of recurrences in conservative systems, however such discovery was not put into practice until the development of faster computers \cite{marwan2007}, for which Eckmann et al. \cite{eckmann1987} in 1987 introduced a method where recurrences in the dynamics of a system can be visualised using
Recurrence Plots. The intention of Eckmann et al. \cite{eckmann1987} was to propose a tool, called Recurrence Plot (RP), that provides insights into high-dimensional dynamical systems where trajectories are very difficult to visualise.
Therefore, "RP helps us to investigate the $m-$dimensional phase space trajectories through a two-dimensional representation of its recurrences" \cite{marwan2015}.
Similarly, Marwan et al. \cite{marwan2015} pointed out that additionally to the methodologies of the state space reconstruction and other dynamic invariants such as Lyapunov exponent, Kolmogorov-Sinai entropy, the recurrences of the trajectories in the phase space can provide important clues to characterise natural process that present, for instance, periodicities (as Milankovitch cycles) or irregular cycles (as El Ni\~no Southern Oscillation).
Such recurrences can not only be presented visually using Recurrence Plots (RP) but also be quantified with Recurrence Quantification metrics, which leads to applications of these in various areas such as economy, physiology, neuroscience, earth science, astrophysics and engineering \cite{marwan2007}.

For the creation of a recurrence plot based on time series $\{ \boldsymbol{x}_n \}$, it is first computed the state space reconstruction with uniform time-delay embedding $X(i)=\{ \boldsymbol{ \tilde{x} }_n, \dots, \boldsymbol{ \tilde{x} }_{n -(m-1)\tau} \}$ where $i=1,\dots,N$, $N$ is the number of considered states of $X(i)$ and $X(i) \in \mathbb{R}^m$ \cite{eckmann1987}.
The recurrence plot is therefore a two-dimensional $N \times N$ square matrix, $\mathbf{R}$, where a black dot is placed at $(i,j)$ whenever $X(i)$ is sufficiently close to $X(j)$:
\begin{equation}
\mathbf{R}^{m}_{i,j} (\epsilon) = \Theta ( \epsilon_i - || X(i) - X(j) ||
\end{equation}
where $\quad i,j=1,\dots,N$, $\epsilon$ is a threshold distance, $|| \cdotp ||$ a norm, and $\Theta(\cdotp)$ is the Heaviside function (i.e. $\Theta(x)=0$, if $x<0$, and $\Theta(x)=1$ otherwise) (Fig~\ref{fig01}(c1,c2)) \cite{eckmann1987, marwan2007,marwan2015}.
RP is also characterised with a line of identity (LOI) which is a diagonal line due to $ R_{i,j}=1 (i,j=1,\dots,N)$.

\subsection*{Structures of Recurrence Plots}
Pattern formations in the RPs can be designated either as topology for large-scale patterns or texture for small-scale patterns.
In the case of topology, the following pattern formations are commonly presented: (i) homogeneous where uniform recurrence points are spread in the RP e.g., uniformly distributed noise (Fig~\ref{fig01}c3), (ii) periodic and quasi-periodic systems where diagonal lines and checkerboard structures represent oscillating systems, e.g., sinusoidal signals (Fig~\ref{fig01}c4), (iii) drift where paling or darkening recurrence points away from the LOI is caused by drifting systems, e.g., logistic map (Fig~\ref{fig01}c5), and (iv) disrupted where recurrence points are presented white areas or bands that indicate abrupt changes in the dynamics, e.g. Brownian motion (Fig~\ref{fig01}c7) \cite{eckmann1987, marwan2015}.
Texture patterns in RPs can be categorised as: (i) single or isolated recurrence points that represent rare occurring states, and do not persist for any time or fluctuate heavily, (ii) dots forming diagonal lines where the length of the small-scale parallel lines in the diagonal are related to the ratio of determinism or predictability in the dynamics of the system, and (iii) dots forming vertical and horizontal lines where the length of the lines represent a time length where a state does not change or change very slowly and these patterns formation represent discontinuities in the signal, and (iv) dots clustering to inscribe rectangular regions which are related to laminar states or singularities \cite{marwan2015}.

Although, each of the previous pattern descriptions of the structures in the RP offer an idea of the characteristics of dynamical systems, these might be misinterpreted and conclusions might tend to be subjective as these require the interpretation of a particular researcher(s).
Because of that, recurrence quantification analyis (RQA) offer objective methodologies to quantify such visual characteristics of previous recurrent pattern structures in the RP \cite{zbilut1992}.

\subsection*{Recurrence Quantifications Analysis (RQA)}
Originally, Zbilut et al. \cite{zbilut1992} proposed metrics to investigate the density of recurrence points in RPs, then histograms of lengths for diagonal lines in RPs were studied by \cite{trulla1996} which were the introduction to the term recurrence quantification analysis (RQA) \cite{marwan2008}.
RQA has been applied in many fields such as life science, engineering, physics, and others \cite{marwan2008}. Particularly in human movement to investigate noise and complexity of postural control \cite{rhea2011}, postural control \cite{apthorp2014} or interpersonal coordination \cite{duran2017}.
The success of RQA is not only due to its simple algorithmic implementation but also to its capacity to detect tiny modulations in frequency or phase which are not detectable using standard methods e.g. spectral or wavelet analysis \cite{marwan2011}, and that RQA's metrics are quantitatively and qualitatively independent of embedding dimension which is verified experimentally by \cite{iwanski1998}.
RQA metrics comprehend percentage of recurrence, percentage of determinism, ratio, Shannon entropy of the frequency distributions of the line lengths, maximal line length and divergence, trend and laminariy \cite{marwan2007, marwan2015}.
For this work, we considered only four RQA metrics, due to its consistency with our preliminary experiments, which are described below.
Such metrics are computed the nonlinearTseries R package \cite{nonlinearTseries2016}.

\subsubsection*{REC values}
The percentage of recurrence (REC) is defined as
\begin{equation}
REC(\epsilon,N) = 
		\frac{1}{N^2 - N} \sum^{N}_{i \neq j = 1} 
		\mathbf{R}^{m}_{i,j}(\epsilon),
\end{equation}
which enumerates the black dots in the RP excluding the line of identity.
REC is a measure of the relative density of recurrence points in the sparse matrix \cite{marwan2015}.

\subsubsection*{DET values} 
The percent determinism (DET) is defined as the fraction of recurrence points that form diagonal lines and it is determined by
\begin{equation}
DET=
	\frac{\sum^{N}_{l=d_{min}} l H_D{l} }
	     {\sum^{N}_{i,j=1} \mathbf{R}_{i,j}(\epsilon) },
\end{equation}
where 
\begin{equation}
H_D(l) = 
	\sum^{N}_{i,j=1} 
	(1- \mathbf{R}_{i-1,j-1}(\epsilon) ) 
	(1- \mathbf{R}_{i+l,j+l}(\epsilon) ) 
	\prod^{l-1}_{k=0}  \mathbf{R}_{i+k,j+k}(\epsilon)
\end{equation}
is the histogram of the lengths of the diagonal structures in the RP.
DET can be interpreted as the predictability of the system for periodic signals which, in essence, have longer diagonal lines than the short diagonals lines for chaotic signals or absent diagonal lines for stochastic signals \cite{marwan2007, marwan2015}.
Similarly, DET is considered as a measurement for the organisation of points in RPs  \cite{iwanski1998}.

\subsubsection*{RATIO values}
RATIO is defined as the ratio between DET and REC and it is calculated from the frequency distributions of the lengths of the diagonal lines.
RATIO is useful to discover dynamic transitions \cite{marwan2015}.
%
%
%

\subsubsection*{ENT values}
ENT is the Shannon entropy of the frequency distribution of the diagonal line lengths and it is defined as
\begin{equation}
ENT= - \sum^{N}_{l=d_{min}} p(l) ln p(l) \quad with 
	\quad p(l)=\frac{ H_D(l) }{ \sum^{N}_{ l=d_{min} } H_D(l) }.
\end{equation}
ENT reflects the complexity of the deterministic structure in the system.
For instance, for uncorrelated noise or oscillations, the value of ENT is rather small and indicates low complexity of the system, therefore "the higher the ENT is the more complex the dynamics are" \cite{marwan2007, marwan2015}.
%
%

\subsection*{Sensitivity and robustness of RPs and RQA.}
RP and RQA are a very young field in nonlinear dynamics and many questions are still open, for instance, different parameters for window length size of the time series, embedding parameters or recurrence threshold can generate different results in RQA's metrics \cite{marwan2011, eckmann1987}.

The selection of recurrence threshold, $\epsilon$, can depend on the system that is analysed. For instance, when studying dynamical invariants $\epsilon$ require to be very small, for trajectory reconstruction $\epsilon$ requires to have a large thresholds or when studying dynamical transition there is little importance about the selection of the threshold \cite{marwan2011}.
Other criteria for the selection of $\epsilon$ is that the recurrence threshold  should be five times larger than the standard deviation of the observational noise or the use of diagonal structures within the RP is suggested in order to find the optimal recurrence threshold for (quasi-)periodic process \cite{marwan2011}.
Similarly, Iwanski et al. \cite{iwanski1998} highlighted the importance of choosing the right embedding parameters to perform RQA for which many experiments have to be performed using different parameters in order to have a better intuition of the nature of the time series and how this is represented by using RQA.

With that in mind, this work explores the sensitivity and robustness of the window size of time series, embedding parameters for RSS with UTDE and recurrence threshold for RP and RQA in order to gain a better insight into the underlying time series collected from inertial sensors in the context of human-humanoid imitation activities.

\section*{Experiment} \label{sec:experiment}
\begin{figure}[ht]
  \centering
\includegraphics[width=1.0\textwidth]{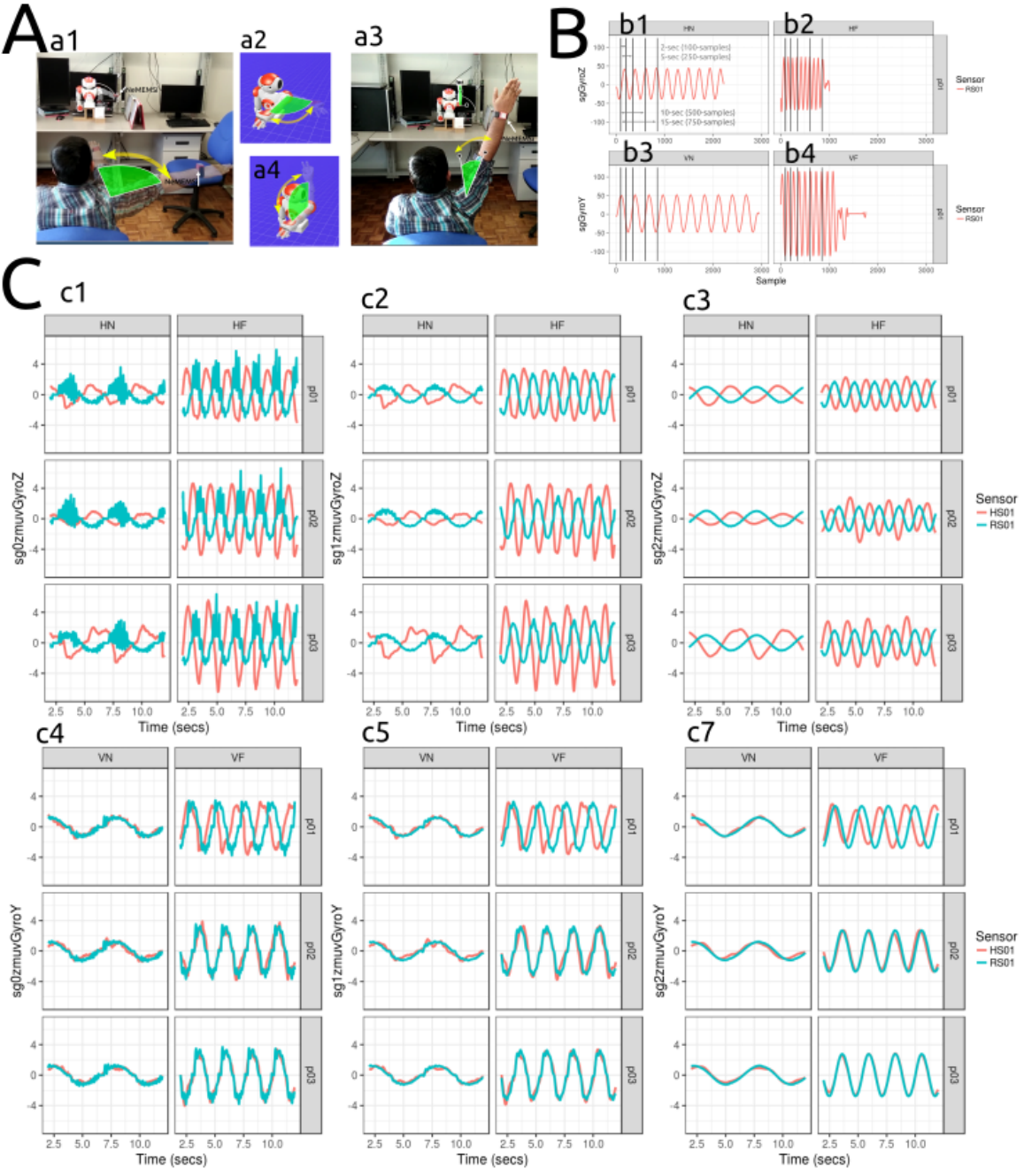}
    \caption{
	{\bf Human-humanoid imitation activities (A). Time series duration for arm movements (B).
		Time series for horizontal and vertical arm movements (C).
	}
	Face-to-face human-humanoid imitation (HHI) activities for (a1) HHI of horizontal arm movement, (a2) Humanoid horizontal arm movement, (a3) HHI of vertical arm movement, and (a4) Humanoid vertical arm movement.
    Time series of smoothed data from gyroscope sensor for different speed arm movements performed by NAO: (b1) Horizontal Normal arm movement (HN), (b2) Horizontal Faster arm movement (HF), (b3) Vertical Normal arm movement (VN) and (b4) Vertical Faster arm movement (VF).
	(B) shows window sizes for 2-seconds (100 samples), 5-seconds (250 samples), 10-seconds (500 samples) and 15-seconds (750 samples).
	(c1,c4) raw-normalised (sg0zmuv), (c2,c5) normalised-smoothed 1 (sg1zmuv) and (c3,c7) normalised-smoothed 2 (sg2zmuv).
	Time series are only for three participants	($p01$, $p02$, and $p03$) for horizontal and vertical arm movements in normal and faster velocity (HN, HF, VN, VF) with the normalised GyroZ or GyroY axis and with one sensor attached to the participant (HS01) and other sensor attached to the robot (RS01).
	Code and data to reproduce the figure is available in \cite{srep2021}.
}
    \label{fig02}
\end{figure}

We conducted an experiment in the context of human-humanoid imitation (HHI) activities where participants were asked to imitate simple horizontal and vertical arm movements performed by NAO, a humanoid robot \cite{gouaillier2009}.
Such simple movements were repeated ten times for the participant who copied NAO's arm movements in a face-to-face imitation activity.
Also, wearable inertial measurement unit (IMU) sensors were attached to the right hand of the participant and to the left hand of the robot (Figure~\ref{fig02}a1,a3).
Data were then collected with four NeMEMSi IMU sensors with sampling rate of 50Hz provinding tri-axial data of the accelerometer, gyroscope and magnetometer sensors and quaternions \cite{Comotti2014}.

\subsection*{Participants}
Twenty-three participants, from now on defined as $pN$ where $N$ is the number of participant, were invited to do the experiment. However, data for three participants were not used because the instructions for $p01$, who was the only left-handed, were mistakenly given in a way that movements were performed different from what had been planned, and for participants $p13$ and $p16$ data were corrupted because of problems with Bluetooth communications with sensors.
With that in mind, data for twenty participants were analysed in this work.
Of the twenty participants, all of them are right-handed healthy participants of whom four are females and sixteen are males with a mean and standard deviation (SD) age of mean=19.8 (SD=1.39).

\subsection*{Ethics}
All participants provided informed consent forms prior to participation in the experiment. Hence, the experiments were conducted in November 2016 and participants confirmed reading and understanding the participant information sheet of the experiments and were able to withdraw from the experiment at any time without giving any reason.
The design of the experiments adhered to the University of Birmingham regulations, data were anonymised and videos were stored only on a personal computer in accordance with the Data Protection Act 1998.

\subsection*{Human-humanoid imitation activities}
For human-humanoid imitation (HHI) activities four neMEMSi sensors were used, two of which were attached to the right hand of the participant and the other two to the left hand of the humanoid robot.
Then, each participant was asked to imitate repetitions of simple horizontal and vertical arm movements performed by the humanoid robot in the following conditions:
(i) ten repetitions of horizontal arm movement at normal (HN) and faster (HF) speed (Figure~\ref{fig02}a1), and
(ii) ten repetitions of vertical arm movement at normal (VN) and faster (VF) speed (Figure~\ref{fig02}a3).
The normal and faster speed of arm movements is defined by the duration in number of samples of one repetition of NAO's arm movements.
We select NAO's arm movements duration to distinguish between normal and faster arm movements as NAO's movements have less variation for such repetitive movements.
The duration for one repetition of the horizontal arm movement at normal speed, HN, is 5 seconds considering that each repetition last around 250 samples.
For horizontal arm movement at faster speed, HF, each repetition were performed in 2 seconds which correspond to 90 samples of data.
The vertical arm movement at normal speed, VN, were performed  in 6 seconds which is around 300 samples of data.
For vertical arm movement at faster speed, VF, each repetition lasts about 2.4 seconds which correspond to 120 samples of data.
To visualise the distinction between normal and faster speed for horizontal and vertical arm movements, Fig~\ref{fig02}B shows smoothed time series for axes Z and Y of the gyroscope sensors with four window lengths: 2-sec (100-samples), 5-sec (250-samples), 10-sec (500-samples) and 15-sec (750-samples).

\subsection*{Data collection from inertial measurement units} 
\label{sec:experiment:subsec:imu}
To give insight to the research questions, we considered various conditions of time series collected for this work:
\begin{itemize}
\item Three levels of smoothness for normalised data (e.g., sg0zmuv, sg1zmuv and sg2zmuv) where sg stands for Savitzky-Golay filter with two different filter lengths (29 and 159) and the same polynomial degree of 5 using the function \texttt{sgolay(p,n,m)} \cite{Rsignal} and zmuv is zero mean unit variance.
\item four arm movement activities with two velocities: horizontal normal (HN), horizontal faster (HF), vertical normal (VN), and vertical faster (VF), and
\item four window lengths: 2-sec (100 samples), 5-sec (250 samples), 10-sec (500 samples) and 15-sec (750 samples).
\end{itemize}
Due to space limitations and to have simple visualisation, we only present 10-sec (500 samples) window length time series for three participants (p01, p01 and p03) performing horizontal arm movements (axis GyroZ) and vertical arm movements (axis GyroY) (Figs \ref{fig02}C).

\subsubsection*{Raw data}
We focus our analysis from data of the accelerometer and gyroscope of the NeMEMsi sensors \cite{Comotti2014} and leave the data of the magnetometer and quaternions for further investigation because of their possible variations with regard to magnetic disturbances \cite{shoaib2016}.
Data from the accelerometer is defined by triaxial time series $A_x(n)$, $A_y(n)$, $A_z(n)$ which forms the matrix $\boldsymbol{A}$ (Eq.~\ref{eq:AG}), and the same for data from the gyroscope which is defined by triaxial time-series of $G_x(n)$, $G_y(n)$, $G_z(n)$ representing the matrix $\boldsymbol{G}$ (Eq.~\ref{eq:AG}).
Both triaxial time series of each sensor, $a$ and $g$, are denoted with its respective axes subscripts $x,y,z$, where $n$ is the sample index and $N$ is the same maximum length of all axes for the time series.
\begin{equation}\label{eq:AG}
\boldsymbol{A} =
\begin{pmatrix}
  A_x(n) \\
  A_y(n) \\
  A_z(n)
\end{pmatrix}
=
\begin{pmatrix}
 a_x(1),a_x(2),\dots,a_x(N) \\
 a_y(1),a_y(2),\dots,a_y(N) \\
 a_z(1),a_z(2),\dots,a_z(N) 
\end{pmatrix},
\boldsymbol{G} =
\begin{pmatrix}
 G_x(n) \\
 G_y(n) \\
 G_z(n)
\end{pmatrix}
=
\begin{pmatrix}
 g_x(1),g_x(2),\dots,g_x(N) \\
 g_y(1),g_y(2),\dots,g_y(N) \\
 g_z(1),g_z(2),\dots,g_z(N) 
\end{pmatrix}.
\end{equation}

\subsubsection*{Postprocessing data}
After the collection of raw data from four NeMEMsi sensors, time synchronisation alignment and interpolation were performed in order to create time series with the same length and synchronised time.
We refer the reader to \cite{Comotti2014} for further details about the time synchronisation process.

\subsubsection*{Normalising data}
Data is normalised to be zero mean and unit variance.
The sample mean and sample standard deviation using $x(n)$ is given by
\begin{equation}\label{eq:ms}
\mu_{x(n)}= \frac{1}{N} ( \sum_{i=1}^N x(i) ), \quad 
	\sigma_{x(n)} =  \sqrt{ \frac{  \sum_{1=1}^N ( x(i) - \mu_{x(n)} )^2 }{ N-1 }  },      
\end{equation}
and the normalised data, $\hat{x}(n)$, is computed as follows
\begin{equation}\label{eq:normalization}
\hat{x} (n) = \frac{   x(n) -  \mu_{x(n)}  }{   \sigma_{x(n)} }.   
\end{equation}

\subsubsection*{Smoothing data}
Commonly, a low-pass filter is use either to capture the low frequencies that represent \%99 of the human body energy or to get the gravitational and body motion components of accelerations \cite{anguita2013}.
However, the elimination of a range of frequencies is not the main focus of this work but the conservation of the structure of time series in terms of their width and heights where, for instance, Savitzky-Golay filter can be a way to accomplish such task \cite{press1992}.
Savitzky-Golay filter is based on the principle of moving window averaging which preserves the area under the curve (the zeroth moment), its mean position in time (the first moment) but the line width (the second moment) is violated and that results, for example, in the case of spectrometric data, into a narrow spectral line with reduced height and width.
With that in mind, the aim of Savitzky-Golay filtering is to find filter coefficients $c_n$ that preserve higher momentums which are based on local least-square polynomial approximations \cite{savitzkygolay1964, press1992, schafer2011}.
Therefore, Savitzky-Golay coefficients are computed with \texttt{sgolay(p,n,m)} in R where \texttt{p} is the filter order, \texttt{n} is the filter length (must be odd) and \texttt{m} is the $m$-th derivative of the filter coefficients \cite{Rsignal}.
Smoothed signal is represented with a tilde over the original variable of the signal: $\tilde{x}(n)$.

\subsubsection*{Windowing data size}
With regards to the window size, Shoaib et al. in 2016 \cite{shoaib2016} investigated effects of seven window lengths (2, 5, 10, 15, 20, 25, 30 seconds) and combination of inertial sensors (accelerometer, gyroscope and linear
acceleration sensor) to improve the activity recognition performance for repetitive activities (walking, jogging and biking) and less repetitive activities (smoking, eating, giving a talk or drinking a coffee).
With that in mind, Shoaib et al. \cite{shoaib2016} concluded that the increase of window length improve the recognition of complex activities because these requires a large window length to learn the repetitive motion patterns.
Particularly, one of the recommendations is to use large window size to recognise less repetitive activities which mainly involve random hand gestures.
Therefore, for the four activities (HN, HF, VN, and VF) in this work, which are mainly repetitive, we select only four window sizes for analysis: 2-s window (100 samples), 5-s window (250 samples), 10-s (500 samples) and 15-s window (750 samples) (Figures~\ref{fig02}B).

\section*{Results}

\subsection*{Reconstructed State Spaces}
As noted in the Introduction, a challenge in the implementation of uniform time-delay embedding arises from the selection of embedding parameters because of the uniqueness of each time series in terms of its structure (e.g., modulation of amplitude, frequency, phase etc.).
With that in mind, the options are to either calculate embedding parameters for each unique instance (which can make comparison challenging) or to find parameters which can apply to all instances in the study.
We recognise that the latter approach is not without its problems, but our approach is to compute sample mean over all values in each of the conditions of the time series for minimum dimension and minimum delay values.

\subsubsection*{Minimum embedding parameters}
\begin{figure}[ht]
\centering
\includegraphics[width=1.0\textwidth]{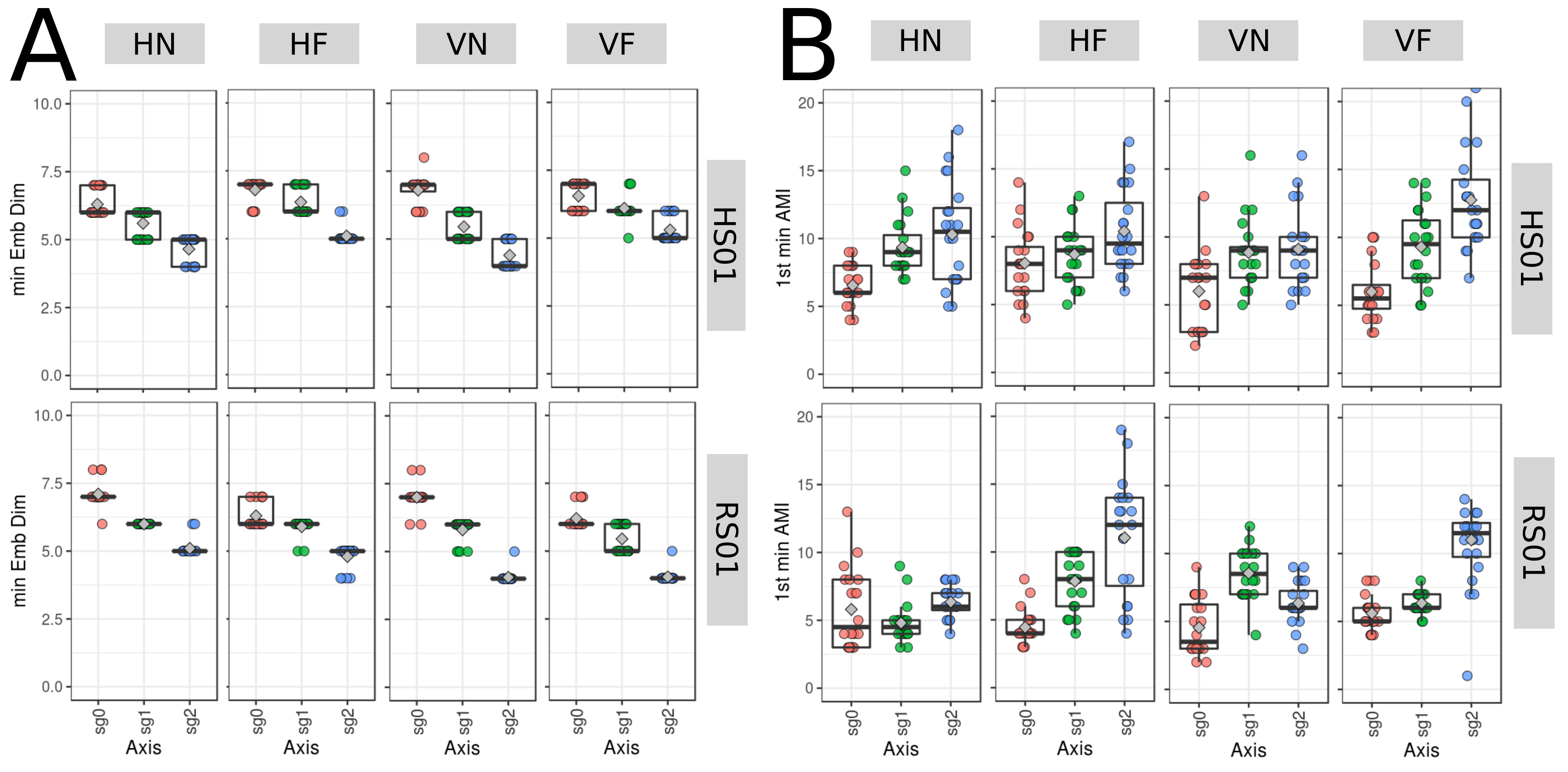}
	\caption{
	{\bf Box plots of minimum embedding parameters.} 
	Box plots of (A) minimum embedding dimensions and (B) first minimum AMI values for Horizontal Normal (HN), Horizontal Faster (HF), Vertical Normal (VN) and Vertical Faster (VF) with sensors attached to participants (HS01) and sensor attached to robot (RS01).
	Minimum embedding parameters are for twenty participants ($p01$ to $p20$) with three smoothed signals (sg0: sg0zmuvGyroZ, sg1: sg1zmuvGyroZ and sg2: sg2zmuvGyroZ) and window length of 10-sec (500 samples).
	Code and data to reproduce the figure is available in \cite{srep2021}.
        }
    \label{fig:cao_ami}
\end{figure}

Minimum embedding parameters were initially computed using False Nearest Neighbour (FNN) and Average Mutual Information (AMI).   We used FNN to calculate a value for $m$ to unfold the attractors and AMI to calculate a value for $\tau$ and maxime the information in the unfolded attractor.
To illustrate this, figure \ref{fig:cao_ami} (A) show box plots for minimum embedding values for sensors on the wrist of the human (HS01) and the robot (RS01).
As we had assumed, minimum embedding values for HS01 show greater variation than those for RS01 (as indicated by differences in interquartile ranges).
Additionally, figure \ref{fig:cao_ami} (A) show a decrease in mean values (rhombus) in the box plots as smoothness of the time-series increases (see sg0, sg1, sg2).
Figure \ref{fig:cao_ami} (B) show box plots for Average Mutual Information (AMI).
One can see that, in contrast to RS01, minimum values for HS01 tend to spread as the smoothness of the time-series increases.
The sample mean for minimum value of embedding parameter, $m$ derived from FNN  (figures \ref{fig:cao_ami} A) is $\overline{m}_0=6$ and $\tau$ from AMI (figures \ref{fig:cao_ami} B) is $\overline{\tau}_0=8$.

\subsubsection*{Uniform Time-Delay Embedding}
\begin{figure}[ht]
\centering
\includegraphics[width=1.0\textwidth]{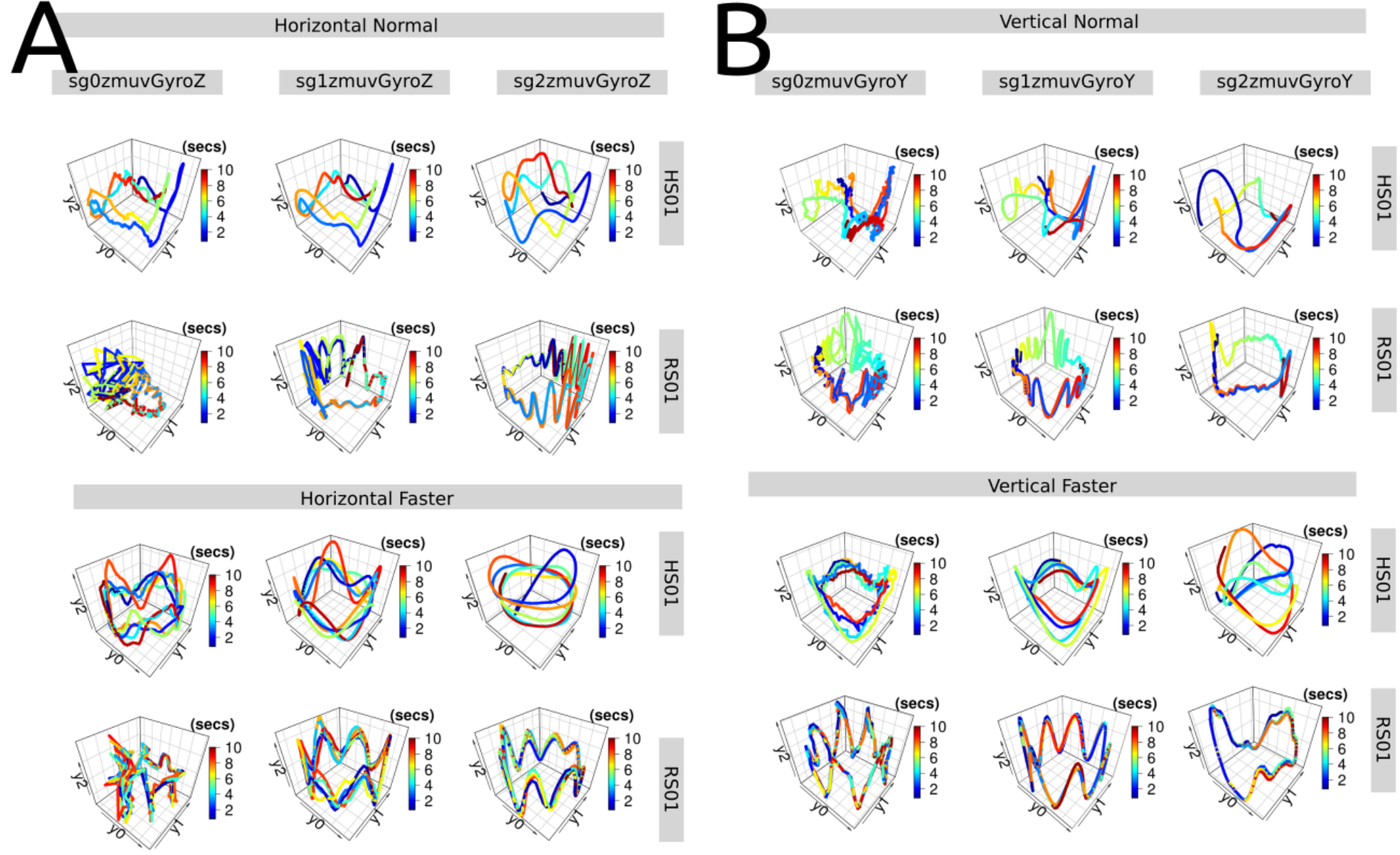}
\caption{
	{\bf RSSs for horizontal and vertical arm movements.}
	Reconstructed state spaces for time series for p01 of Figure \ref{fig02}C.
	Reconstructed state spaces were computed with embedding parameters $\overline{m}_0=6$, $\overline{\tau}_0=8$
	for (A) horizontal and (B) vertical arm movements.
	Code and data to reproduce the figure is available in \cite{srep2021}.	
        }
    \label{fig04}
\end{figure}

Using the overall embedding parameters ($\overline{m_0}=6$, $\overline{\tau_0}=8$), the first three axis of the rotated Principal Components Analysis (PCA) are shown for the reconstructed state spaces of horizontal (figures~\ref{fig04}(A)) and vertical (figures~\ref{fig04}(B)) arm movements.
While visual inspection of these figures suggests differences in the trajectories of the reconstructed state spaces, we require an objective quantification to determine the extent of the differences.
One approach could be use Euclidean distances from the origin for points in these trajectories, but this proved inconclusive and was not able to capture the dynamics of the trajectories.
Consequently, we applied Recurrence Plots and Recurrence Quantification Analysis.

\subsection*{Recurrences Plots}
\begin{figure}[ht]
\centering
\includegraphics[width=1.0\textwidth]{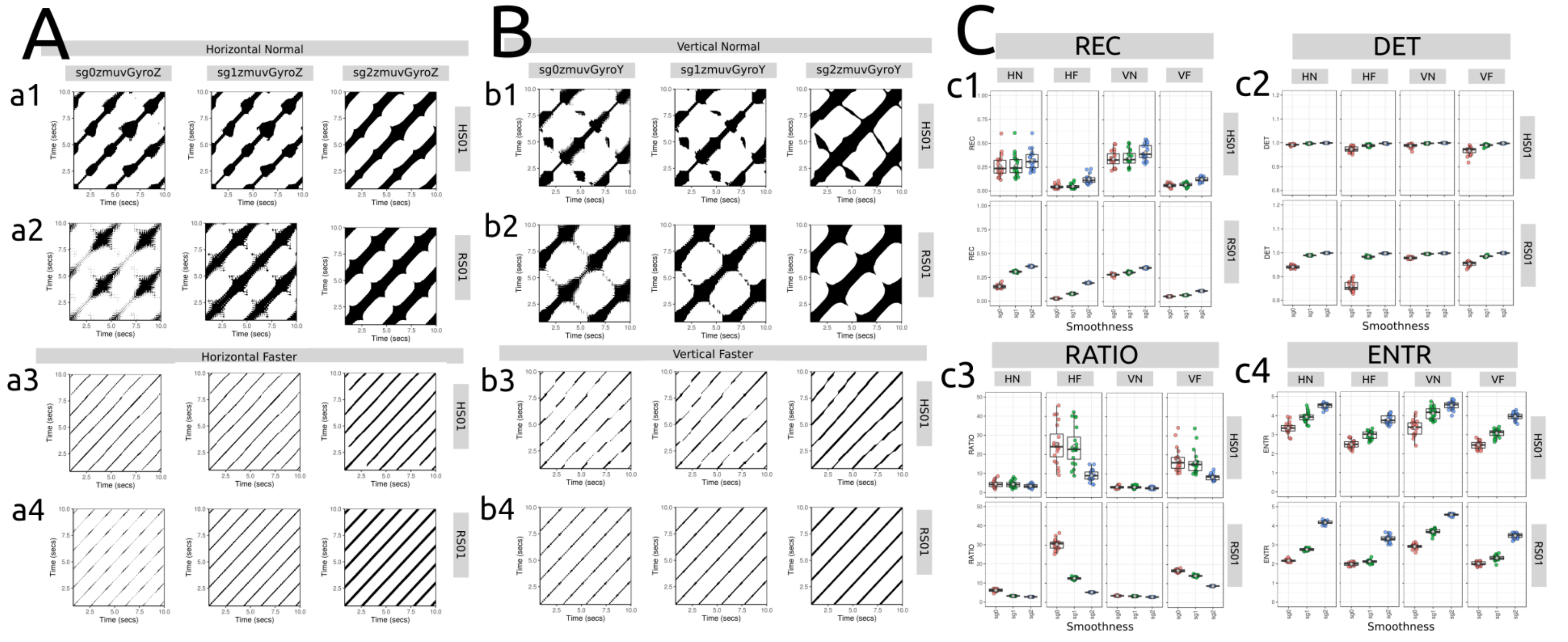}
\caption{
	{\bf RPs for horizontal (A) and vertical (B) arm movements.}	
	Recurrence plots were computed with embedding parameters
	$\overline{m}_0=6$, $\overline{\tau}_0=8$, and $\epsilon=1$.
	{\bf Box plots for RQA metrics (C).}
	RQA metrics for (c1) REC, (c2) DET, (c3) RATIO, and (c4) ENTR of 20 participants performing HN, HF, VN and VF movements with sensors HS01, RS01 and three smoothed-normalised time series (sg0, sg1 and sg2).
	RQA values were computed with $\overline{m}_0=6$, $\overline{\tau}_0=8$, and $\epsilon=1$.
	Code and data to reproduce the figure is available in \cite{srep2021}.
        }
    \label{fig05}
\end{figure}

Using the average embedding parameters ($\overline{m}_0=6$, $\overline{\tau}_0=8$) and an recurrence threshold of $\epsilon=1$.
As our interest is for dynamical transitions, there is little importance on the selection of $\epsilon$ which in this case is 1.
Recurrence Plots (RP) were computed for horizontal (figures~\ref{fig05}(A)) and vertical (figures~\ref{fig05}(B)) arm movements.

\subsection*{Recurrence Quantification Analysis} \label{ch6:rqas}
Four Recurrence Quantification Analysis metrics (percentage recurrence, REC, representing the percentage of black dots in RP; percentage determinism, DET, representing the predictability of the RP; ratio of DET / REC, RATIO; Shannon entropy, ENTR) were computed using the same parameters as for RP.
Figure~\ref{fig05}(c1) presents box plots of REC values, for HS01, are more spread for Slow (i.e., 5 seconds per movement) movements in Horizontal (HS) or Vertical (VS) than for Fast (i.e., 2 seconds per movement).
This suggests greater variation between participants for the Slow movements.  
For RS01, there is little variation between Slow and Fast movement (interquartile range of 0.01).
In terms of smoothness, there seems little effect of HS01 but RS01 values do show affects of smoothness (see the incremental changes of mean values (rhombus)).
Figure \ref{fig05}(c2) presents DET values and shows little difference for type of movement or performer.
However, DET values are affected by changes in smoothness of the signal, particularly for Fast movement.
Figure \ref{fig05}(c3) presents the ratio of DET / REC.
These values, for the human performer, vary less for HN than for HF movement.  
Additionally, smoothness leads to a decrease in mean values for Fast movements.
Figure \ref{fig05}(c4) shows ENTR values are higher for the human
performer than the robot, and vary with the smoothness of the time-series.

\subsection*{3D RQA ENTR}
\begin{figure}
\centering
\includegraphics[width=1.0\textwidth]{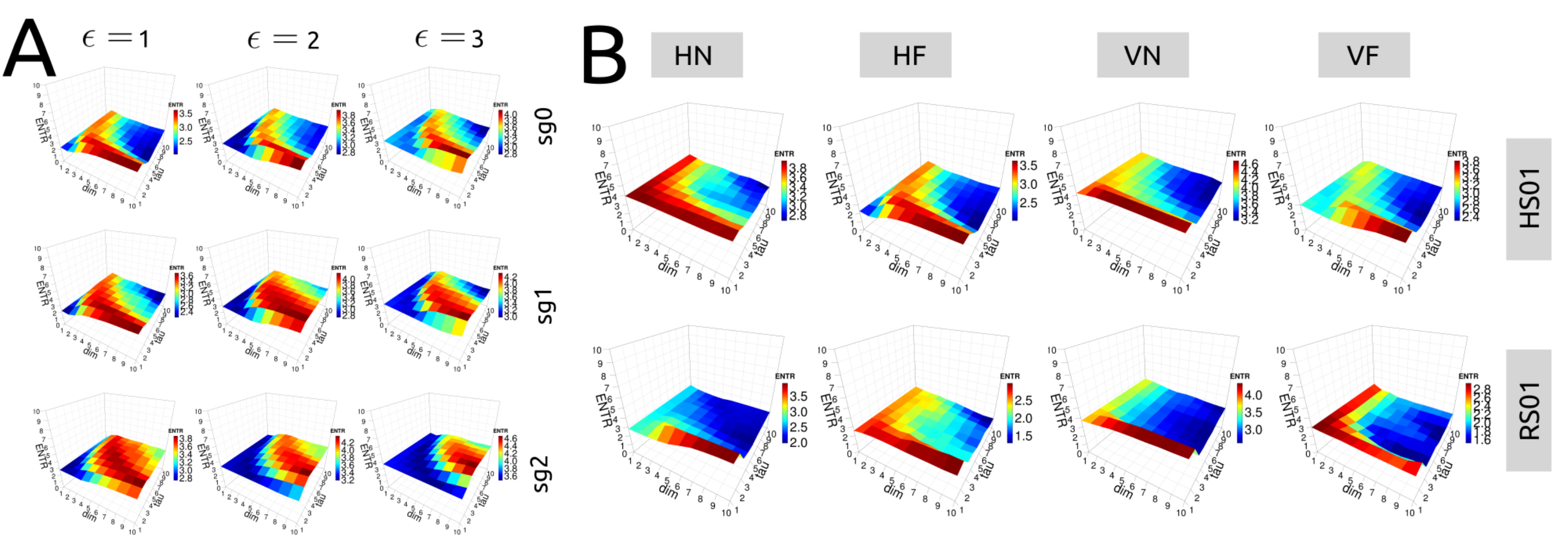}
    \caption
	[3D surface plots of RQA ENTR values]{
	{\bf 
	3D surface plots of RQA ENTR values for different recurrence threshold and smoothness levels (A) and for different sensors and activities (B).}
	RQA ENTR values are for embedding parameters $ \{ m \in \mathbb{R} | 0 \le m \le 10  \} $, $ \{ \tau \in \mathbb{R} | 0 \le \tau \le 10  \}$ incrementing by one and three recurrence thresholds $\epsilon=1, 2, 3$.
	RQA ENTR values were for A were computed with data from $p03$, sensor HS01, with a window size of 10-secs (500 samples).
	RQA ENTR values are for embedding parameters $ \{ m \in \mathbb{R} | 0 \le m \le 10  \}$, $ \{ \tau \in \mathbb{R} | 0 \le \tau \le 10  \}$ with $\epsilon = 1 $ considering four activities: Horizontal Normal (HN), Horizontal Faster(HF), Vertical Normal(VN), Vertical Faster (VF), and sensors: Human Sensor 01 (HS01) and Robot Sensor (RS01).
	RQA ENTR values for B were computed from data of $p03$, sg0 and window size of 10-secs (500 samples).
	Code and data to reproduce the figure is available in \cite{srep2021}.
        }
    \label{fig06}
\end{figure}

As ENTR appeared to be have higher sensitivity than the other measures, we explored the impact of different embedding parameters $ \{ m \in \mathbb{R} | 1 \le m \le 10  \} $, $  \{ \tau \in \mathbb{R} | 1 \le \tau \le 10  \} $ incrementing by 1 each run, with recurrence thresholds $\epsilon=1, 2, 3$ and levels of smoothness (sg0, sg1, sg2).
Figure \ref{fig06}(A) shows that increasing recurrence threshold leads to an increase in ENTR regardless of level of smoothness.
Similarly, increasing level of smoothness will also increase ENTR (figure \ref{fig06}(A)).
In terms of movement or performer, RQA ENTR decrease from Slow (HS, VS) to Fast (HF, VF) for both human (HS01) and (RS01) (Figures~\ref{fig06}(B)).
In terms of individual differences, for human participants, figure \ref{fig:3dRQAENTR_participantsactivities} compares p01, p02 and p03 (for illustrative purposes) -- both between each other and in comparison with the more consistent performance of the robot.
In terms of window length (w100 (2s), w250 (5s), w500(10s), w750 (15s)), figure \ref{fig:3dRQAENTR_windowsactivities} shows that improvement in capture with number of samples, although this has less of an effect on RQA ENTR.

\begin{figure}[ht]
\centering
\includegraphics[width=0.9\textwidth]{fig07}
    \caption{
	{\bf 3D surface plots of RQA ENTR values for different participants, activities and sensors.}
	RQA ENTR values are for participants (p01, p02, and p03) 
	in the categories of 
	(A) Human Sensor 01 (HS01) and 
	(B) Robot Sensor 01 (RS01)
	considering embedding parameters
	$ \{ m \in \mathbb{R} | 0 \le m \le 10  \}$,
	$ \{ \tau \in \mathbb{R} | 0 \le \tau \le 10  \}$
	with $\epsilon = 1$ and four activities 
	Horizontal Normal (HN), Horizontal Faster(HF), Vertical Normal(VN), and 
	Vertical Faster (VF).
	RQA ENTR values were computed from data of sg0 and window size of 10-secs (500 samples).
	Code and data to reproduce the figure is available in \cite{srep2021}.
       }
\label{fig:3dRQAENTR_participantsactivities}
\end{figure}

\begin{figure}[ht]
\centering
\includegraphics[width=1.0\textwidth]{fig08}
    \caption{
	{\bf 3D surface plots of RQA ENTR values for different windows lengths and activities.}
	RQA ENTR values are for embedding parameters
	$ \{ m \in \mathbb{R} | 0 \le m \le 10  \}$,
	$ \{ \tau \in \mathbb{R} | 0 \le \tau \le 10  \}$, 
	with $\epsilon = 1 $ considering four 
	windows lengths (e.g., w100 (100 samples), w250 (250 samples),
	w500 (500 samples) and w750 (750 samples)) and
	four activities 
	Horizontal Normal (HN), Horizontal Faster(HF), Vertical Normal(VN), and 
	Vertical Faster (VF).
	RQA ENTR values were computed from data of $p01$ and sg0.
	Code and data to reproduce the figure is available in \cite{srep2021}.
       }
\label{fig:3dRQAENTR_windowsactivities}
\end{figure}

To summarise this section of results, it can be said that computing embedding parameters for individual structure of time-series data is already a solved problem \cite{frank2010, sama2013, bradley2015}.
However, it has been shown the challenge of finding embedding parameters for nonlinear dynamic tools that represent a set of different time-series data.
That said, we proposed the use of sample mean of the set of embedding parameters for RSSs, RP and RQA to then noticed that the selection of recurrence threshold, $\epsilon$, is also an open problem.
For which, this work proposed the variation of recurrence thresholds and embedding parameters to show the relationships of these to different datasets (participants, activities, windows lengths and sensors).

\section*{Discussion}
While there are many approaches to estimating embedding parameters for nonlinear analysis, these can be influenced by the structure of the time-series data.
We show that, for RSS, RP and RQA, the estimation of embedding parameters can be performed using a sample mean which, together with recurrence threshold, can be shown to be influenced by activity, performer, window length and smoothness of time-series.
It is known that time-series from different sources and with different characteristics require different embedding parameters, and this can produce different RSS, RPs and RQAs.
Although this work helps to understand the open problem of finding right balance among (i) the level of smoothness of the signal, (ii) the selection of recurrence thresholds and (iii) the range of embedding parameters, we have shown how RQA metrics can help to quantify movement variability.

\section*{Conclusions}
In this paper we show how the selection of nonlinear analysis tool (i.e., RSS, RP, RQA metrics) depends on what question one wishes to address with time-series data (e.g., predictability, organisation, dynamics, complexity).
Time-series data characteristics (e.g., window length, smoothness), time-series structure (e.g., frequency, amplitude) and data source (e.g., sensor placement, performance, movement) all influence the results that nonlinear analysis methods can provide.
That said, it has been shown that the use of different characteristics of the data and their collection can help us visualise and quantify variability of movement using methods of nonlinear analysis.
There remain limitations of nonlinear methods in relation to the estimation of parameters (e.g., recurrence threshold, embedding parameters) which reflect the dynamics of specific movement and performers, window length  and structure of the time-series.
We note that RQA DET seems to show low sensitivity to these differences, whereas REC and RATIO (primarily as a result of REC) show variation across performers and movements.
RQA ENTR, with different recurrence thresholds, can quantify variation in the time-series data and offers the most appropriate means for analysing variability in movement to allow us to analyse individual differences between human performers.
To then found out that RQA ENTR values with different recurrence thresholds were appropriate to quantify the different changes and variations of the characteristics of time-series data.
Therefore, the use of Shannon Entropy could be applied to analyse human participants who might vary in age, state of health, anthropometric features and capability to perform movement.



\section*{Acknowledgements}
This work was funded by the University of Birmingham and the Mexican National Council of Science and Technology as part of Miguel Xochicale's PhD degree.
I would like to thank to Dr. Dolores Columba Perez Flores and Prof. Martin J Russell for their helpful comments to polish the use of the language of mathematics and to Constantino Antonio Garcia Martinez et al. for developing the R package nonlinearTseries that help to accelerate the analysis of the nonlinear time series in this work.

\section*{Author contributions statement}
Contributions for this work of Miguel Xochicale (MX) and Chris Baber (CB) are as follows:
\begin{description}
\item[Conceptualisation] MX, CB
\item[Data Curation] MX
\item[Formal Analysis] MX
\item[Funding Acquisition] MX, CB
\item[Investigation] MX
\item[Methodology] MX
\item[Project Administration] MX
\item[Resources] CB
\item[Software] MX
\item[Supervision] CB
\item[Validation] MX
\item[Verification] MX
\item[Writing - Original Draft Preparation] MX
\item[Writing - Review] MX, CB
\item[Writing - Editing] MX
\end{description}


\end{document}


\maketitle

\begin{abstract}
Report for supplementary material where section 1 describes datasets and section 2 shows surface plots of 3D RQA ENTR values for 3 participants (p01, p02 and p03).
\end{abstract}

\tableofcontents

\section{Datasets}
Datasets are for 
(a) window size of 100, 250, 500 and 750 samples (w100, w250, w500, w750);
(b) smoothness: sg0, sg1 and sg2;
(c) movement: Horizontal Normal (HN), Horizontal Faster (HF), Vertical Normal (VN) and Vertical Faster (VF); and 
(d) sensors: Sensor attached to Human (HS01), sensor attached to Robot (RS01).

The location of the dataset is shown below with the first two lines of xdatav00.dt denoting labels.
\begin{verbatim}
~/srep2021/data/dataset$ tree --si
[ 46M]  xdata_v00.dt
.
.
.
"Participant" "Activity" "Sensor" "Sample" "Time" 
"sg0zmuvGyroY" "sg1zmuvGyroY" "sg2zmuvGyroY" 
"sg0zmuvGyroZ" "sg1zmuvGyroZ" "sg2zmuvGyroZ"
...
"p01" "HN" "HS01" 1 0 
0.0110396263359954 0.00606430191548277 0.0385586376765087 
0.00467559072085554 0.00509907428630649 0.168927517477539
...
\end{verbatim}


\section{3D RQA ENTR values}
Location of code, data and figures for 3D RQA ENTR values is shown below 
\begin{verbatim}
## Code
$HOME/srep2021/code/rscripts/G_3Drqa
`> source(  paste( getwd(), '/C_3Drqa_plots_epsilons.R', sep=''), echo=TRUE )`

## Data
$HOME/srep2021/data/rqa$ tree -s 

## Figures
$HOME/srep2021/docs/figures/rqa/src/3drqa_epsilons$ tree -s
\end{verbatim}

For the following plots, datasets for NN participants where NN is 01, 02 and 03 are:
\begin{verbatim}
RQAs_pNNw100.dt
RQAs_pNNw250.dt
RQAs_pNNw500.dt
RQAs_pNNw750.dt
\end{verbatim}

\subsection{Participant 01}

\begin{figure}[ht!]
\centering
\includegraphics[scale=1.0]{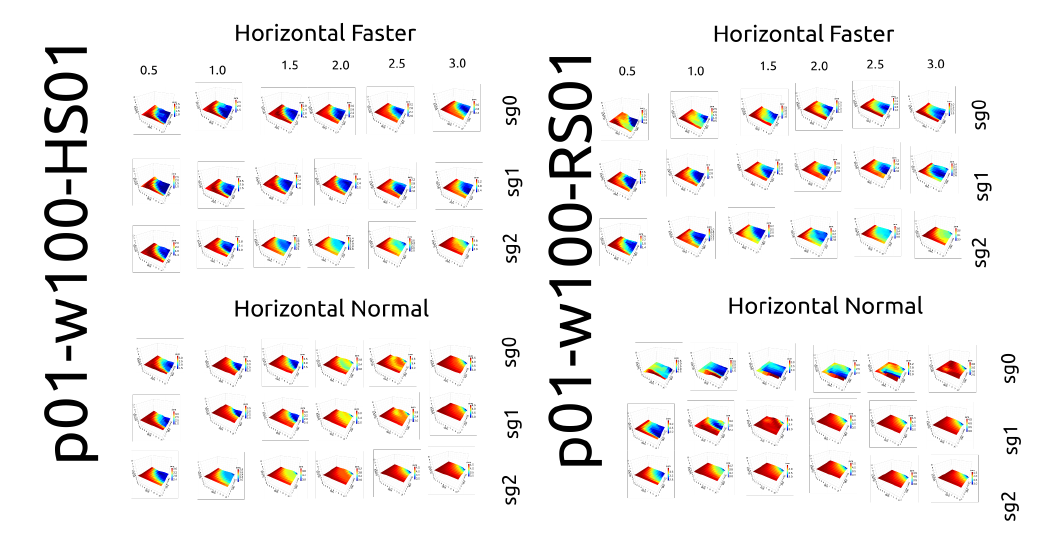}
    	\caption{
	{\bf RQA-Entr for participant 01 performing horizontal movements for a window size of 100 samples.}
	Code and data to reproduce the figure is available in \cite{srep2021}.
        }
    \label{fig-p01-H-w100}
\end{figure}

\begin{figure}[hb!]
\centering
\includegraphics[scale=1.0]{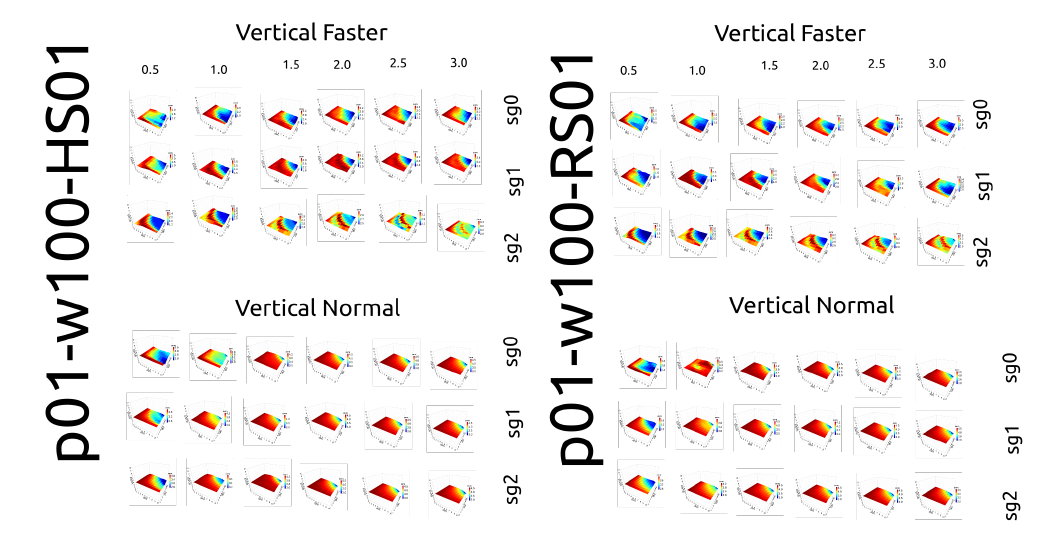}
    	\caption{
	{\bf RQA-Entr for participant 01 performing vertical movements for a window size of 100 samples.}
	Code and data to reproduce the figure is available in \cite{srep2021}.
        }
    \label{fig-p01-V-w100}
\end{figure}

\begin{figure}[ht!]
\centering
\includegraphics{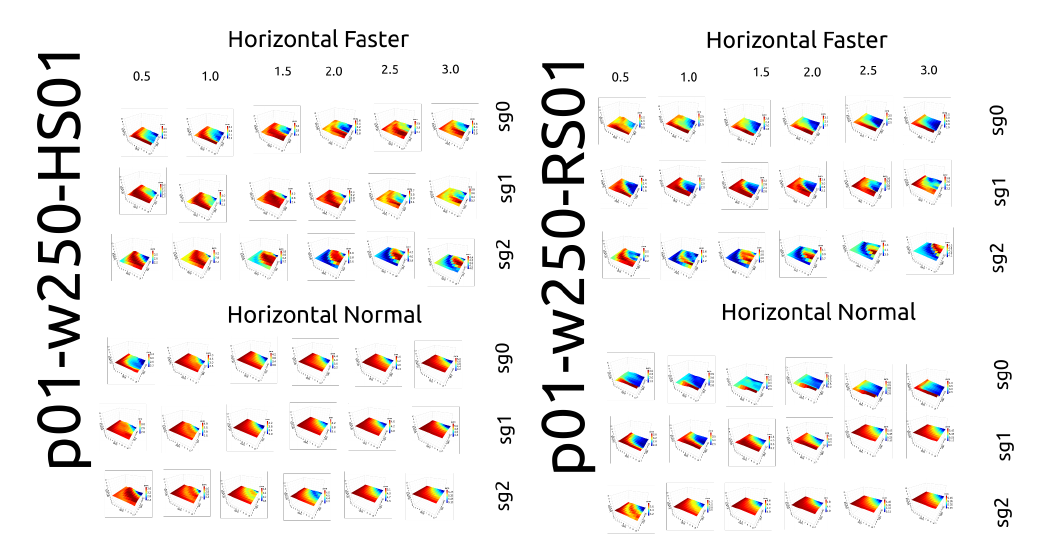}
    	\caption{
	{\bf RQA-Entr for participant 01 performing horizontal movements for a window size of 250 samples.}
	Code and data to reproduce the figure is available in \cite{srep2021}.
        }
    \label{fig-p01-H-w250}
\end{figure}

\begin{figure}[hb!]
\centering
\includegraphics{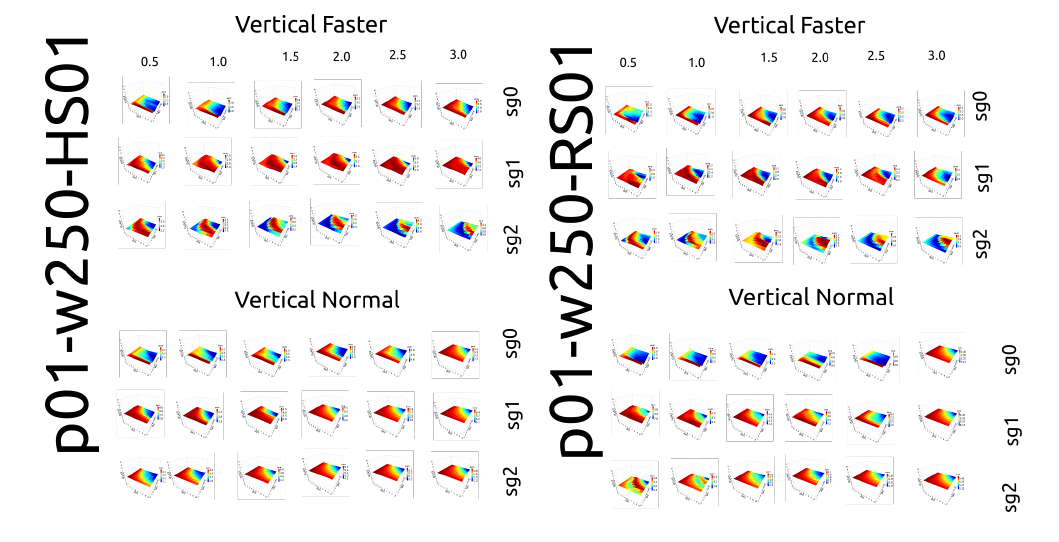}
    	\caption{
	{\bf RQA-Entr for participant 01 performing vertical movements for a window size of 250 samples.}
	Code and data to reproduce the figure is available in \cite{srep2021}.
        }
    \label{fig-p01-V-w250}
\end{figure}

\begin{figure}[ht!]
\centering
\includegraphics{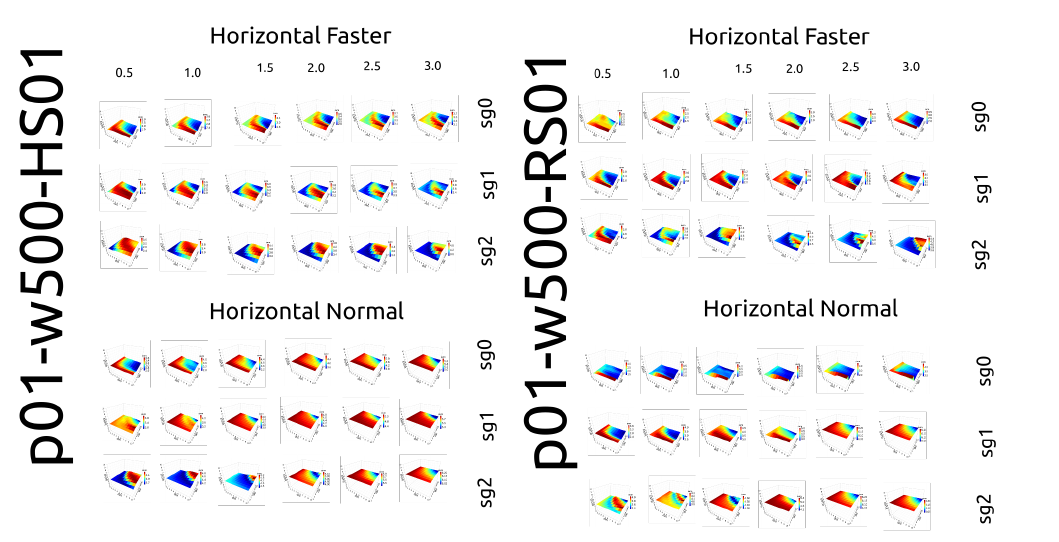}
    	\caption{
	{\bf RQA-Entr for participant 01 performing horizontal movements for a window size of 500 samples.}
	Code and data to reproduce the figure is available in \cite{srep2021}.
        }
    \label{fig-p01-H-w500}
\end{figure}

\begin{figure}[hb!]
\centering
\includegraphics{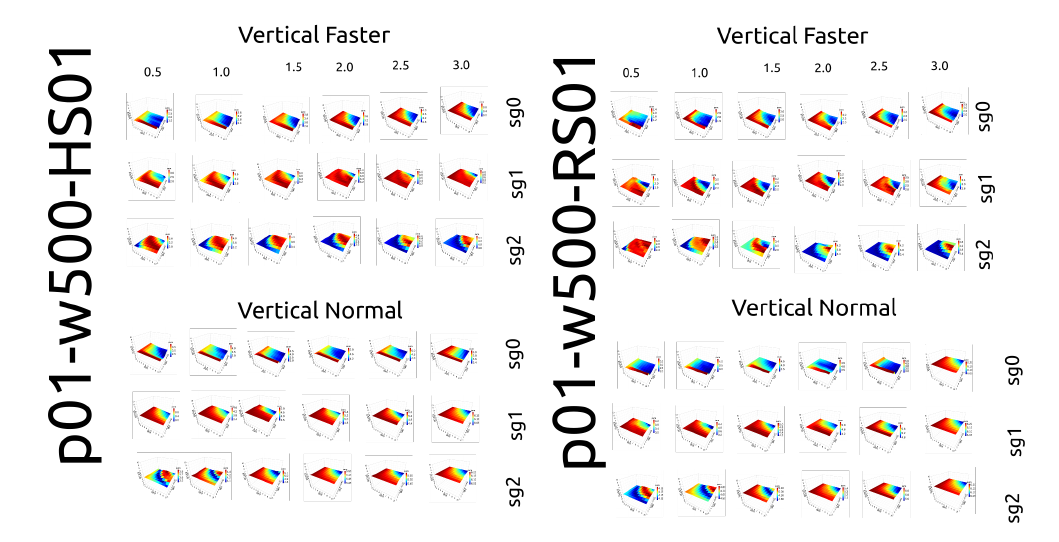}
    	\caption{
	{\bf RQA-Entr for participant 01 performing vertical movements for a window size of 500 samples.}
	Code and data to reproduce the figure is available in \cite{srep2021}.
        }
    \label{fig-p01-V-w500}
\end{figure}

\begin{figure}[ht!]
\centering
\includegraphics{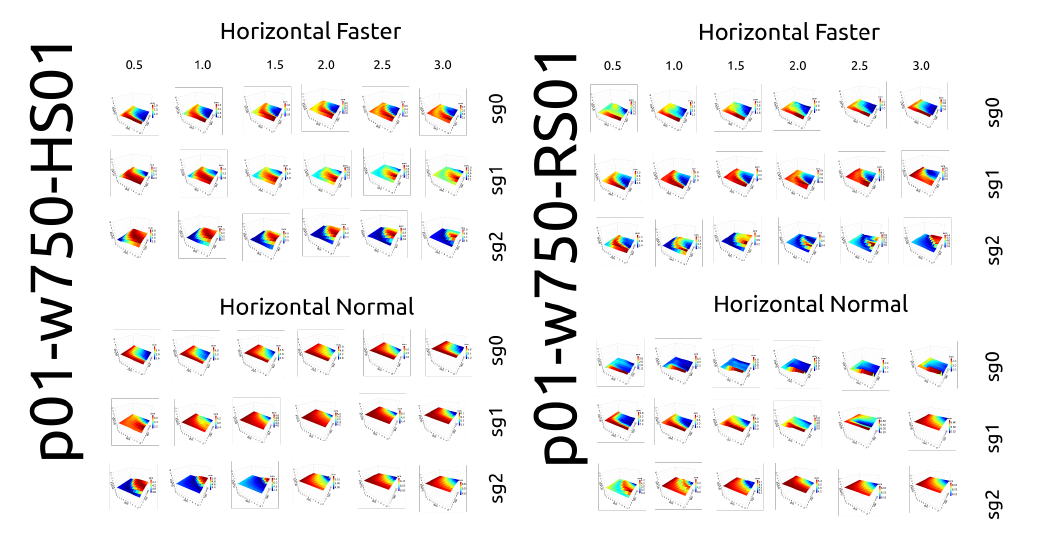}
    	\caption{
	{\bf RQA-Entr for participant 01 performing horizontal movements for a window size of 750 samples.}
	Code and data to reproduce the figure is available in \cite{srep2021}.
        }
    \label{fig-p01-H-w750}
\end{figure}

\begin{figure}[hb!]
\centering
\includegraphics{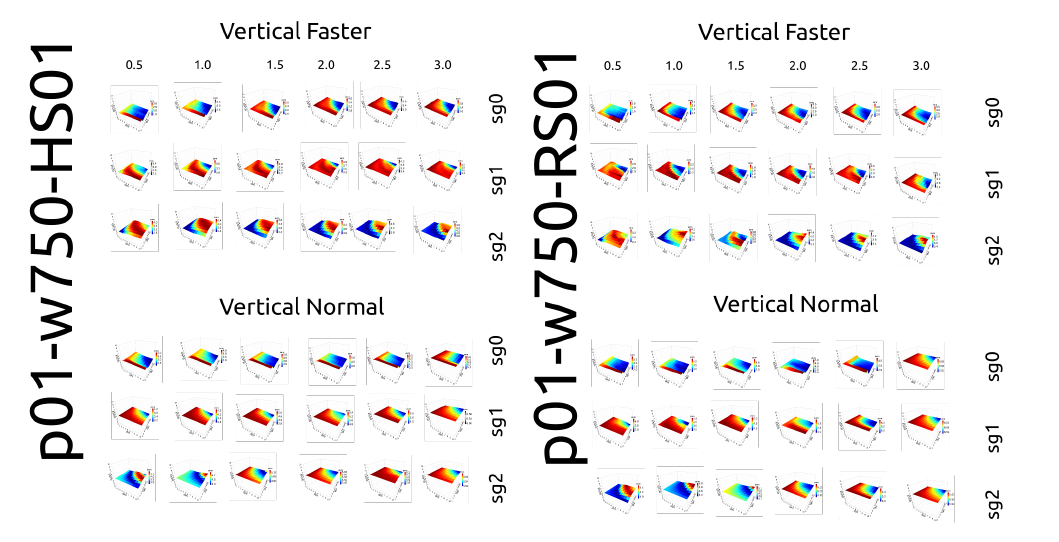}
    	\caption{
	{\bf RQA-Entr for participant 01 performing vertical movements for a window size of 750 samples.}
	Code and data to reproduce the figure is available in \cite{srep2021}.
        }
    \label{fig-p01-V-w750}
\end{figure}

Figures \ref{fig-p01-H-w100} and \ref{fig-p01-V-w100} are for a window size of 100 samples.
Figures \ref{fig-p01-H-w250} and \ref{fig-p01-V-w250} are for a window size of 250 samples.
Figures \ref{fig-p01-H-w500} and \ref{fig-p01-V-w500} are for a window size of 500 samples.
Figures \ref{fig-p01-H-w750} and \ref{fig-p01-V-w750} are for a window size of 750 samples.


\newpage

\newpage

\newpage

\newpage

\newpage
\subsection{Participant 02}
\begin{figure}[ht!]
\centering
\includegraphics[scale=1.0]{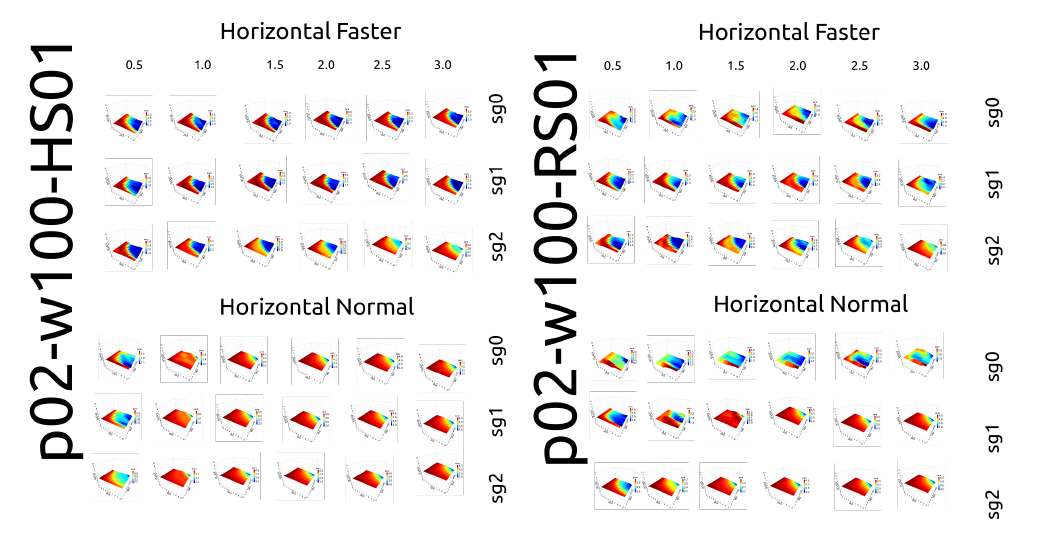}
    	\caption{
	{\bf RQA-Entr for participant 02 performing horizontal movements for a window size of 100 samples.}
	Code and data to reproduce the figure is available in \cite{srep2021}.
        }
    \label{fig-p02-H-w100}
\end{figure}

\begin{figure}[hb!]
\centering
\includegraphics[scale=1.0]{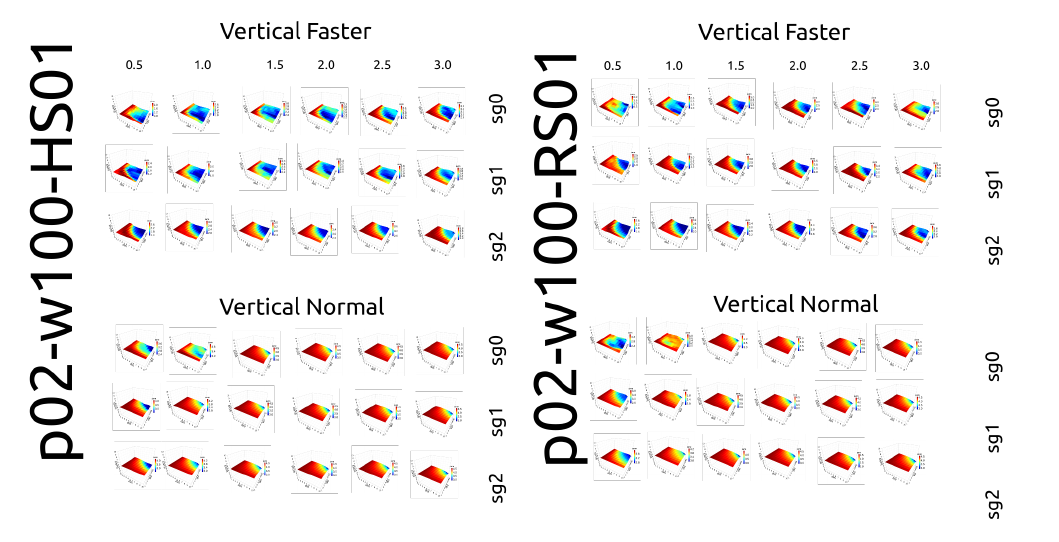}
    	\caption{
	{\bf RQA-Entr for participant 02 performing vertical movements for a window size of 100 samples.}
	Code and data to reproduce the figure is available in \cite{srep2021}.
        }
    \label{fig-p02-V-w100}
\end{figure}

\begin{figure}[ht!]
\centering
\includegraphics{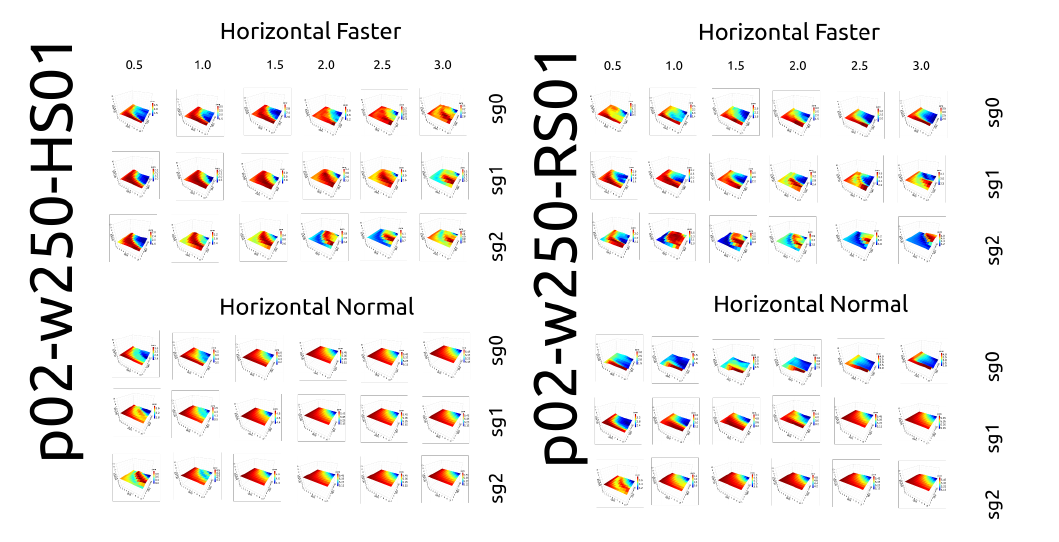}
    	\caption{
	{\bf RQA-Entr for participant 02 performing horizontal movements for a window size of 250 samples.}
	Code and data to reproduce the figure is available in \cite{srep2021}.
        }
    \label{fig-p02-H-w250}
\end{figure}

\begin{figure}[hb!]
\centering
\includegraphics{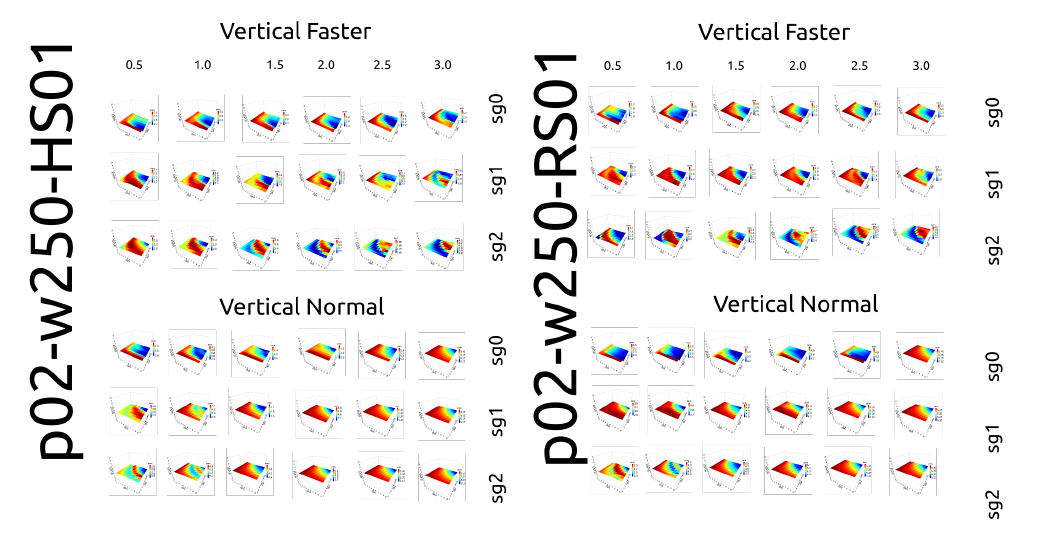}
    	\caption{
	{\bf RQA-Entr for participant 02 performing vertical movements for a window size of 250 samples.}
	Code and data to reproduce the figure is available in \cite{srep2021}.
        }
    \label{fig-p02-V-w250}
\end{figure}

\begin{figure}[ht!]
\centering
\includegraphics{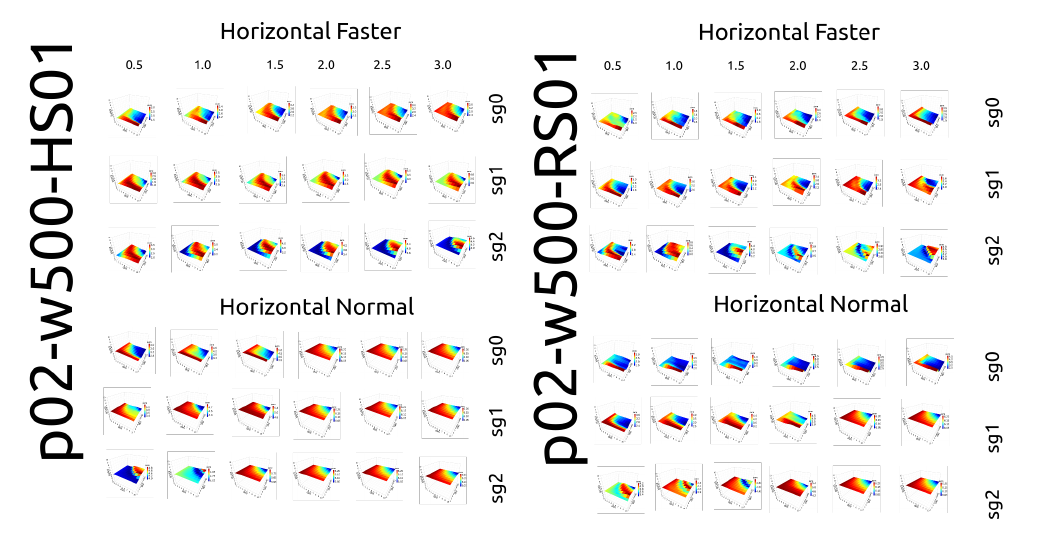}
    	\caption{
	{\bf RQA-Entr for participant 02 performing horizontal movements for a window size of 500 samples.}
	Code and data to reproduce the figure is available in \cite{srep2021}.
        }
    \label{fig-p02-H-w500}
\end{figure}

\begin{figure}[hb!]
\centering
\includegraphics{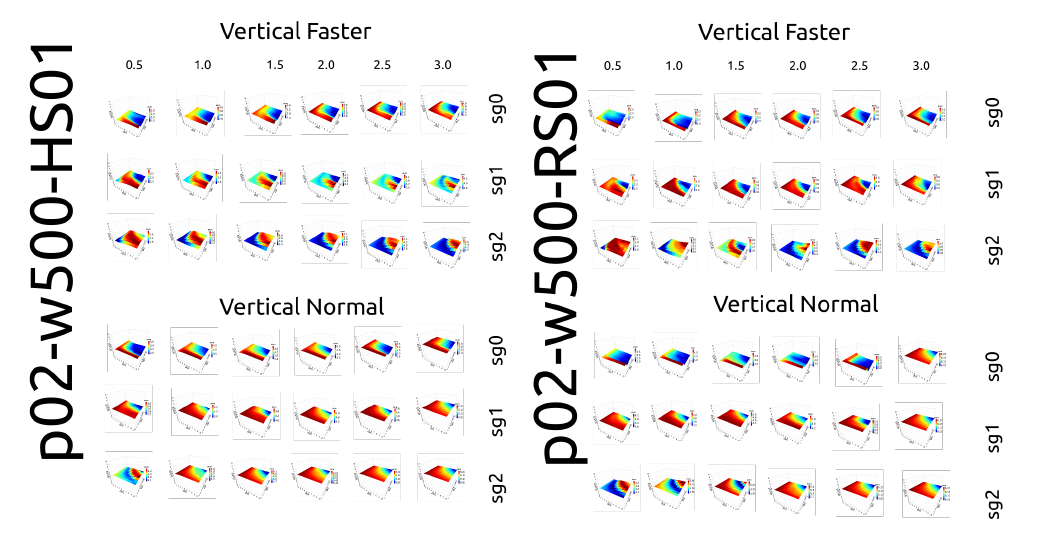}
    	\caption{
	{\bf RQA-Entr for participant 02 performing vertical movements for a window size of 500 samples.}
	Code and data to reproduce the figure is available in \cite{srep2021}.
        }
    \label{fig-p02-V-w500}
\end{figure}

\begin{figure}[ht!]
\centering
\includegraphics{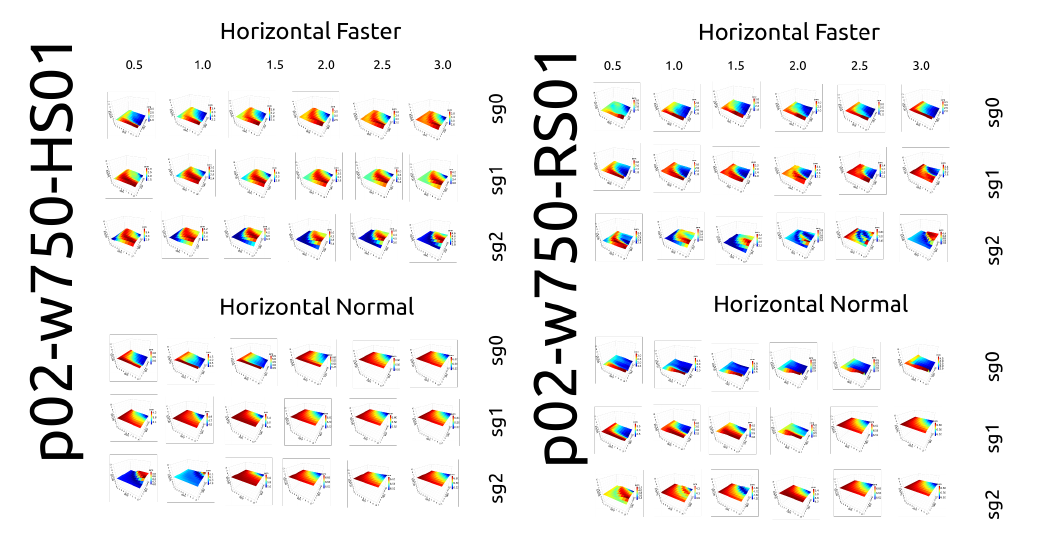}
    	\caption{
	{\bf RQA-Entr for participant 02 performing horizontal movements for a window size of 750 samples.}
	Code and data to reproduce the figure is available in \cite{srep2021}.
        }
    \label{fig-p02-H-w750}
\end{figure}

\begin{figure}[hb!]
\centering
\includegraphics{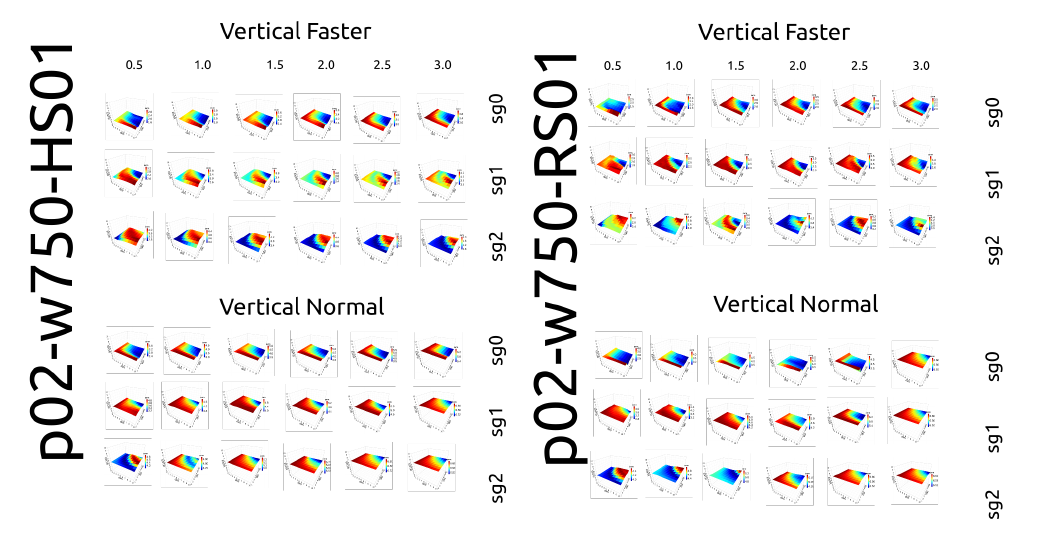}
    	\caption{
	{\bf RQA-Entr for participant 02 performing vertical movements for a window size of 750 samples.}
	Code and data to reproduce the figure is available in \cite{srep2021}.
        }
    \label{fig-p02-V-w750}
\end{figure}

Figures \ref{fig-p02-H-w100} and \ref{fig-p02-V-w100} are for a window size of 100 samples.
Figures \ref{fig-p02-H-w250} and \ref{fig-p02-V-w250} are for a window size of 250 samples.
Figures \ref{fig-p02-H-w500} and \ref{fig-p02-V-w500} are for a window size of 500 samples.
Figures \ref{fig-p02-H-w750} and \ref{fig-p02-V-w750} are for a window size of 750 samples.

\newpage

\newpage

\newpage

\newpage

\newpage
\subsection{Participant 03}
\begin{figure}[ht!]
\centering
\includegraphics[scale=1.0]{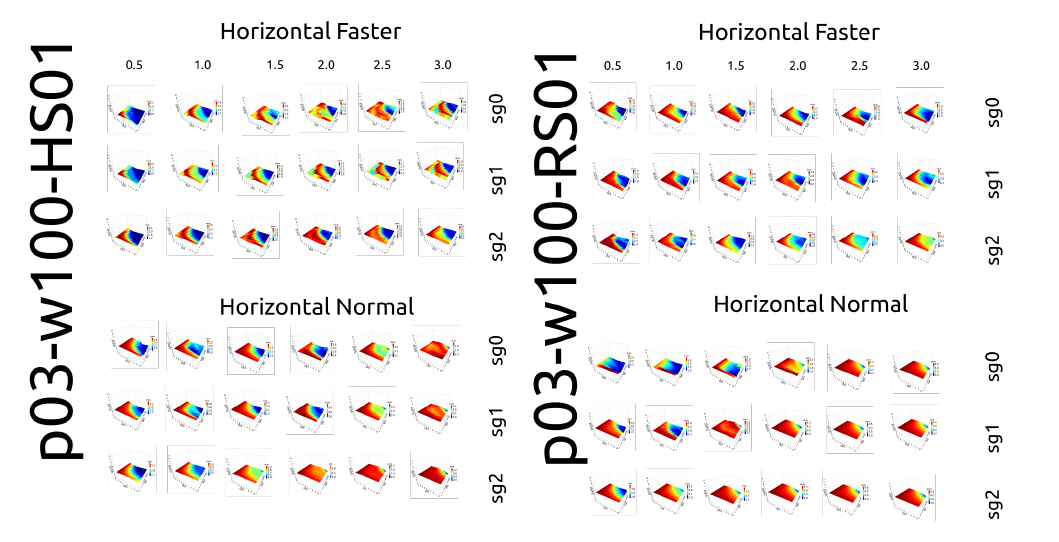}
    	\caption{
	{\bf RQA-Entr for participant 03 performing horizontal movements for a window size of 100 samples.}
	Code and data to reproduce the figure is available in \cite{srep2021}.
        }
    \label{fig-p03-H-w100}
\end{figure}

\begin{figure}[hb!]
\centering
\includegraphics[scale=1.0]{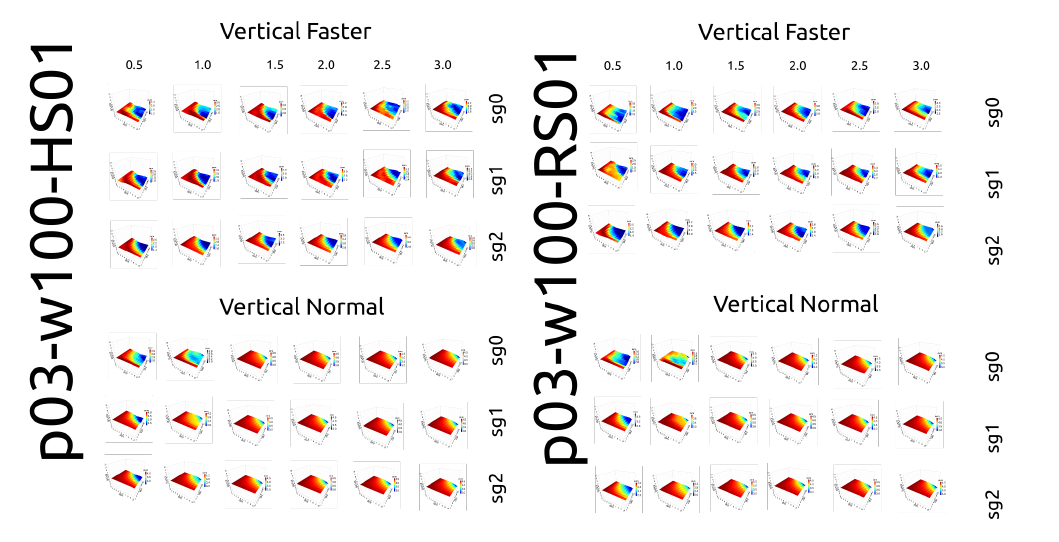}
    	\caption{
	{\bf RQA-Entr for participant 03 performing vertical movements for a window size of 100 samples.}
	Code and data to reproduce the figure is available in \cite{srep2021}.
        }
    \label{fig-p03-V-w100}
\end{figure}

\begin{figure}[ht!]
\centering
\includegraphics{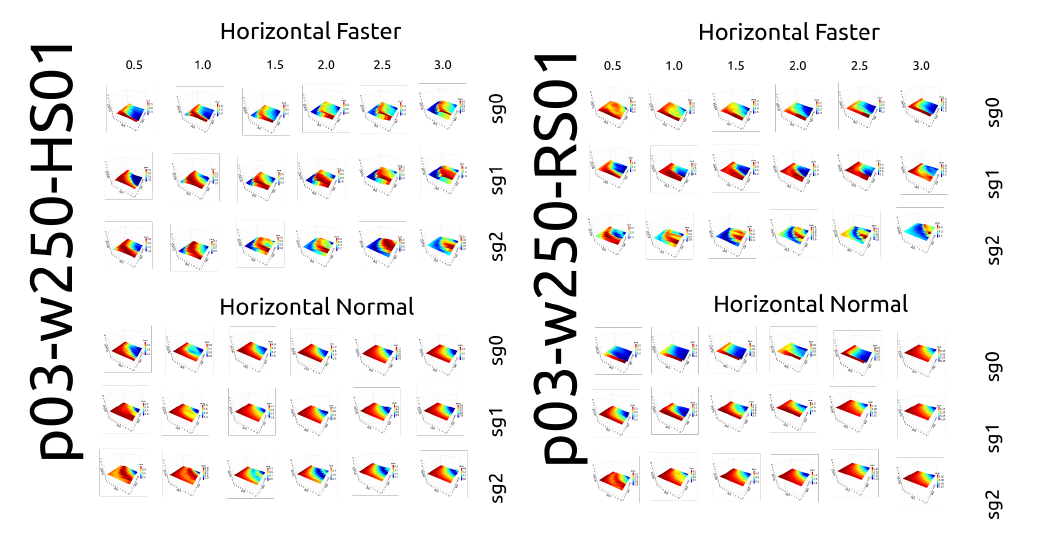}
    	\caption{
	{\bf RQA-Entr for participant 03 performing horizontal movements for a window size of 250 samples.}
	Code and data to reproduce the figure is available in \cite{srep2021}.
        }
    \label{fig-p03-H-w250}
\end{figure}

\begin{figure}[hb!]
\centering
\includegraphics{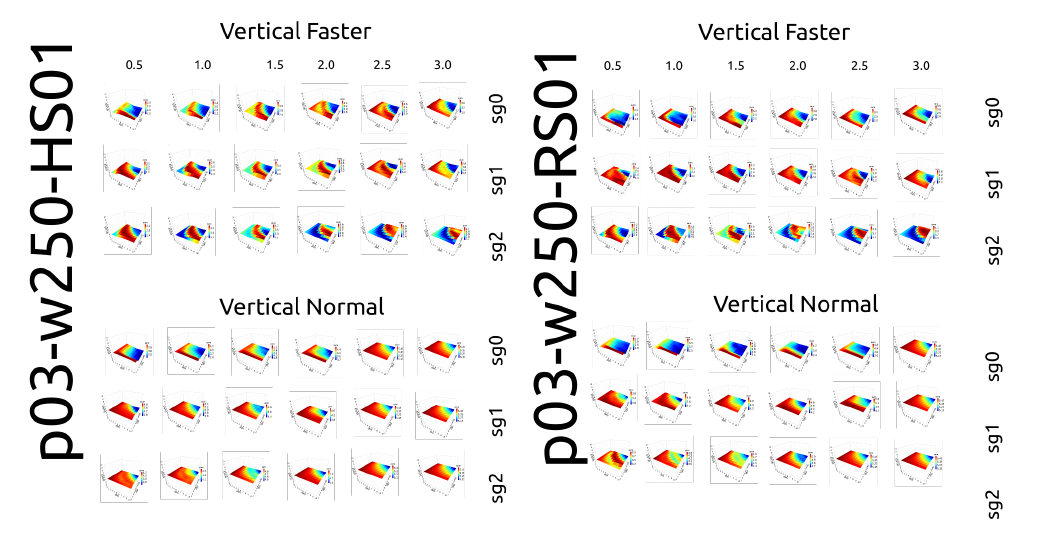}
    	\caption{
	{\bf RQA-Entr for participant 03 performing vertical movements for a window size of 250 samples.}
	Code and data to reproduce the figure is available in \cite{srep2021}.
        }
    \label{fig-p03-V-w250}
\end{figure}

\begin{figure}[ht!]
\centering
\includegraphics{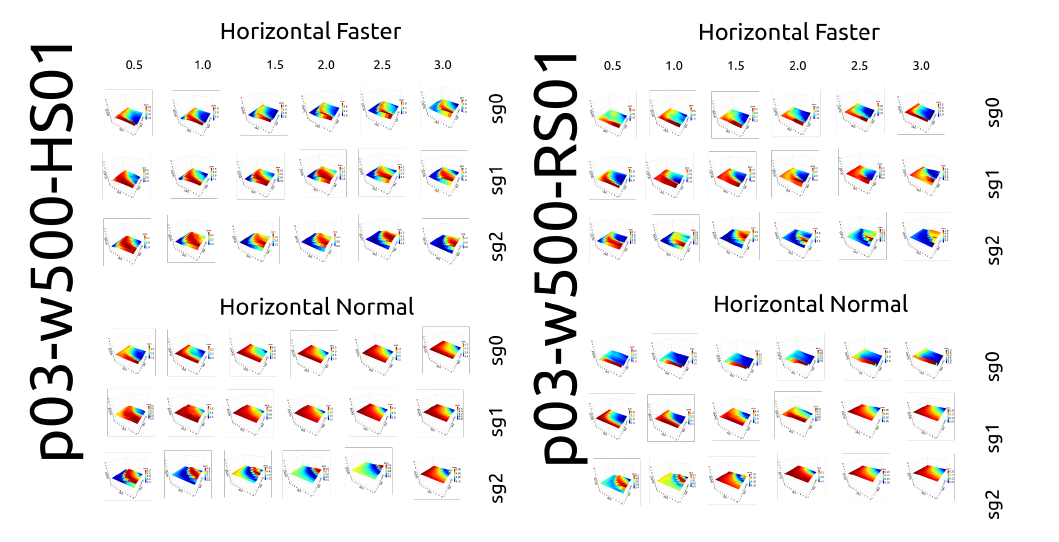}
    	\caption{
	{\bf RQA-Entr for participant 03 performing horizontal movements for a window size of 500 samples.}
	Code and data to reproduce the figure is available in \cite{srep2021}.
        }
    \label{fig-p03-H-w500}
\end{figure}

\begin{figure}[hb!]
\centering
\includegraphics{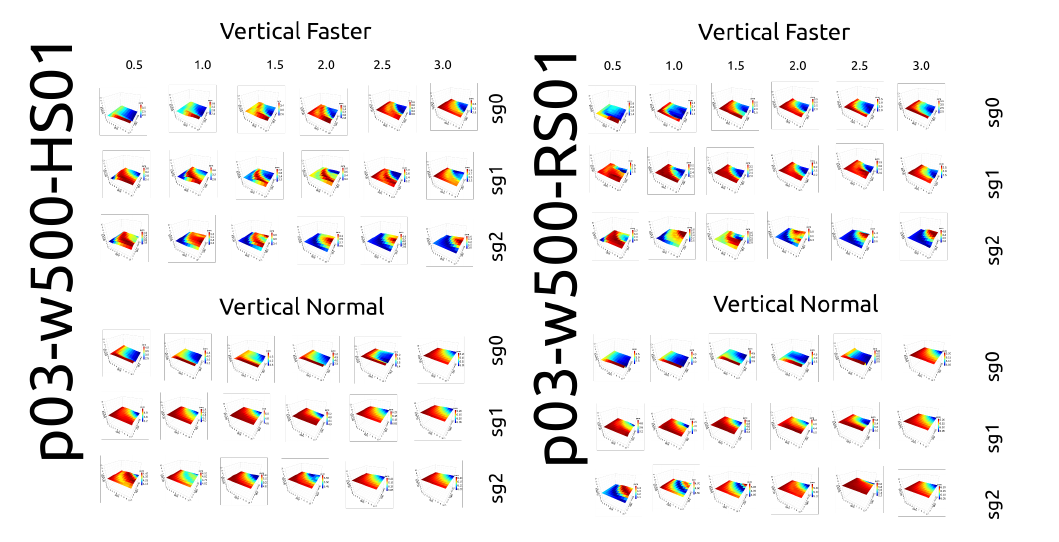}
    	\caption{
	{\bf RQA-Entr for participant 03 performing vertical movements for a window size of 500 samples.}
	Code and data to reproduce the figure is available in \cite{srep2021}.
        }
    \label{fig-p03-V-w500}
\end{figure}

\begin{figure}[ht!]
\centering
\includegraphics{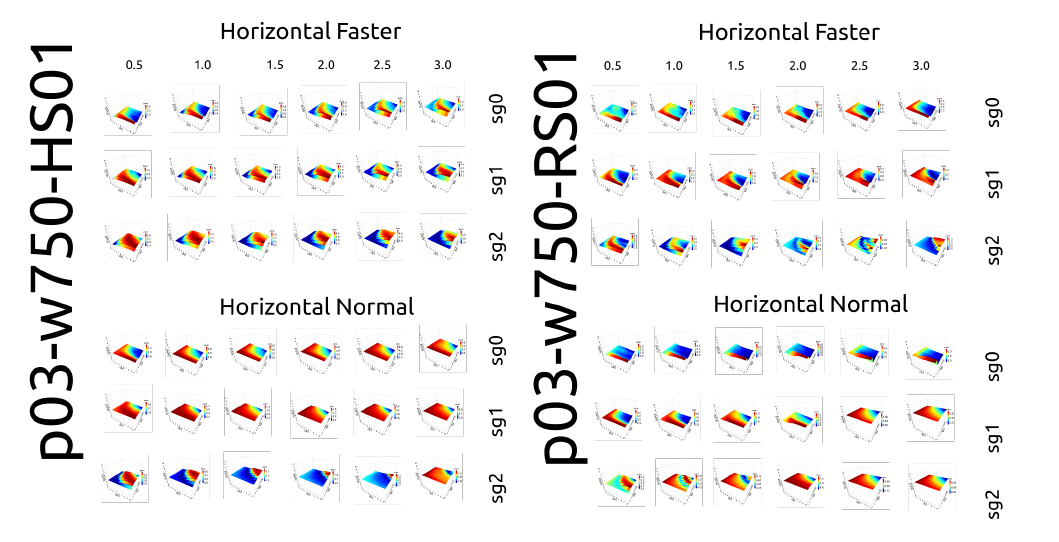}
    	\caption{
	{\bf RQA-Entr for participant 03 performing horizontal movements for a window size of 750 samples.}
	Code and data to reproduce the figure is available in \cite{srep2021}.
        }
    \label{fig-p03-H-w750}
\end{figure}

\begin{figure}[hb!]
\centering
\includegraphics{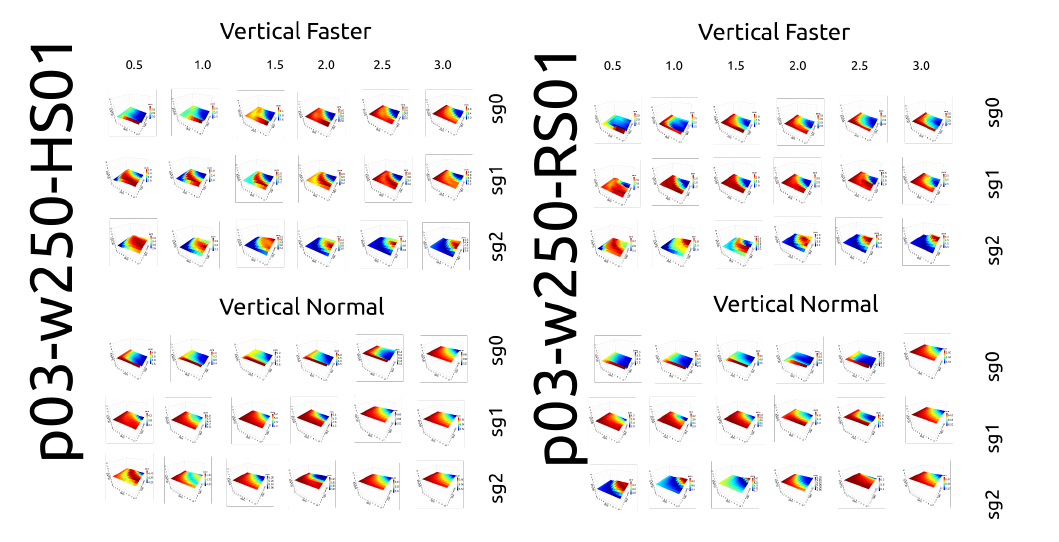}
    	\caption{
	{\bf RQA-Entr for participant 03 performing vertical movements for a window size of 750 samples.}
	Code and data to reproduce the figure is available in \cite{srep2021}.
        }
    \label{fig-p03-V-w750}
\end{figure}

Figures \ref{fig-p03-H-w100} and \ref{fig-p03-V-w100} are for a window size of 100 samples.
Figures \ref{fig-p03-H-w250} and \ref{fig-p03-V-w250} are for a window size of 250 samples.
Figures \ref{fig-p03-H-w500} and \ref{fig-p03-V-w500} are for a window size of 500 samples.
Figures \ref{fig-p03-H-w750} and \ref{fig-p03-V-w750} are for a window size of 750 samples.

\newpage

\newpage

\newpage

\newpage

%
%

\clearpage
\bibliographystyle{plain}
